\documentclass[journal]{IEEEtran}
\usepackage{times,amsmath,amssymb,amsfonts,graphicx,epsfig,cite,psfrag}
\usepackage{algorithmicx,algorithm}
\usepackage{color}

\def\pibf{\boldsymbol \pi }

\def\Thetabf{\boldsymbol \Theta}

\def\abf{{\bf a}}

\def\Oc{{\cal O}}

\def\defeq{~{\stackrel{\Delta}{=}}~}
\def\eg{{\it e.g.,\ \/}}

\def\ie{{\it i.e.,\ \/}}
\def\nn{\nonumber}

\newtheorem{theorem}{Theorem}
\newtheorem{lemma}{Lemma}
\newtheorem{proposition}{Proposition}

\begin{document}

\title{Real-Time Residential-Side Joint Energy Storage Management and Load Scheduling with Renewable Integration}
\author{\IEEEauthorblockN {Tianyi Li, \textit{Student Member, IEEE}, and Min Dong, \textit{Senior Member, IEEE}
\thanks{This work was supported by the National Sciences and Engineering Research Council (NSERC) of Canada under Discovery Grant RGPIN-2014-05181.}
\thanks{A preliminary version of this work \cite{Li&Dong:ICASSP2016} was presented at the 41st \emph{IEEE International Conference on Acoustic Speech and Signal Processing} (ICASSP), Shanghai, China, March 2016.}
\thanks{The authors are with Department of Electrical, Computer, and Software Engineering, University of Ontario Institute of Technology, Oshawa, Ontario, Canada L1H 7K4 (email: tianyi.li@uoit.ca,\ min.dong@uoit.ca).}
}}
\maketitle

\begin{abstract}
We consider joint  energy storage management and load scheduling at a residential site with integrated renewable generation.
Assuming unknown arbitrary dynamics of renewable source, loads, and
electricity price, we aim at optimizing the load scheduling and energy storage control simultaneously in order to minimize the overall system cost within a finite time period.
Besides  incorporating battery operational constraints and costs, we model each individual load task by its requested power intensity and service durations, as well as the maximum and average delay requirements. To tackle this finite time horizon stochastic problem, we propose a real-time scheduling and storage control solution by applying a sequence of modification and transformation to employ Lyapunov optimization  that otherwise is not directly applicable.
With our proposed algorithm, we show that the joint load scheduling and energy storage control can in fact be separated and sequentially determined. Furthermore, both scheduling and energy control decisions have closed-form solutions for simple implementation. Through analysis, we show that our proposed real-time algorithm has a bounded performance guarantee from the optimal $T$-slot look-ahead solution and is asymptotically equivalent to it as the battery capacity and time period goes to infinity.
The effectiveness of joint  load scheduling and energy storage control  by our proposed algorithm is demonstrated through simulation as compared with
alternative algorithms.
\end{abstract}


\section{Introduction}
\label{sec:Introduction}
The rising global demand of energy has resulted in high prices for electricity and also caused the growing environmental concern due to excess carbon emission from power generation. Integrating renewable energy sources into the grid system has become a vital green energy solution to reduce the energy cost and  build a sustainable society and economy. Although promising, renewable energy is often intermittent and difficult to predict, making it less reliable for both grid-level operation and as a local energy source for consumers. Energy storage and flexible loads are considered as two promising management solutions to mitigate the randomness of renewable generation, as well as to reduce electricity cost  \cite{CastilloGayme:Elsevier14,Callaway&Hiskens:ProcIEEE11}. In particular, energy storage can be exploited to shift energy across time, while flexible loads can be controlled to shift demand across time. For grid operators, they can be utilized to counter the fluctuation in renewable generation and to  increase reliability. For consumers, energy storage and load scheduling can provide effective means for energy management to reduce electricity cost.

As renewable penetration into the power supply increases, the renewable generation with storage solutions at residential homes (such as roof-top solar panels) will become increasingly popular. Thus, developing a cost effective energy storage management system to maximally harness energy from renewable sources is of critical importance. At the same time, many smart appliances have been developed, creating more controllable loads at the consumer side. They can be controlled to benefit from the dynamic price set at the utility, and help shift the energy demand from high-peak to low-peak periods  to reduce energy bills. Providing effective management solution that combines both energy storage and load scheduling will be the most promising future solution for consumers to reduce energy costs and is the goal of this paper.

Developing an effective joint energy storage management and load scheduling solution is important, but faces unique challenges. For energy storage, the cost reduction by storage comes with an additional cost from battery degradation due to charging and discharging; finite battery  capacity
makes the storage control decisions coupled over time which are difficult to
optimize.
For load scheduling, while minimizing the electricity cost, it needs to ensure the delay requirements for each load and for the overall service are met.
In particular,
load scheduling decision affects the energy usage and storage and vise versa. Thus, storage control and load scheduling decisions are coupled with each other and over time, making it especially challenging for a joint design.

Energy storage management alone has been considered for power balancing to counter  the fluctuation of renewable generation and increase grid reliability \cite{SuGamal:TPS13,SunDongLiang:TSG14,SunDongLiang:JSTSP14}, and for consumers to reduce electricity costs \cite{Wang&etal:TSG14,Lin&etal:NET13,Zhang&Schaar:JTSP14,Urgaonkar&Neely:SIGMETRICS2011,Huang&Walrand&Ramchandran:SmatGridComm12,
Sergio&Ming&Pan&Yong_TSG13,Li&Dong:ICASSP2013,Li&Dong:JSAC15}.
Off-line storage control strategies for dynamic systems have been proposed \cite{SuGamal:TPS13,Wang&etal:TSG14,Lin&etal:NET13}.
In these works, renewable energy arrivals are assumed known ahead of time and the knowledge of load statistics is assumed.
For real-time storage management design,
\cite{Zhang&Schaar:JTSP14} has formulated the storage management control as a Markov Decision Process (MDP) and solved it by Dynamic Programming (DP).
Lyapunov optimization technique \cite{book:Neely} has  been recently employed for designing real-time storage control  at either grid operator side or consumer side under different system models and optimization goals  \cite{Urgaonkar&Neely:SIGMETRICS2011,Huang&Walrand&Ramchandran:SmatGridComm12,Sergio&Ming&Pan&Yong_TSG13,
Li&Dong:ICASSP2013,Li&Dong:JSAC15,SunDongLiang:TSG14,SunDongLiang:JSTSP14}.
Among these works, \cite{Huang&Walrand&Ramchandran:SmatGridComm12,Sergio&Ming&Pan&Yong_TSG13} have considered renewable generation without  modeling the battery operational cost.
Both renewable generation and battery operation cost have been modeled in \cite{SunDongLiang:TSG14,SunDongLiang:JSTSP14,Li&Dong:ICASSP2013,Li&Dong:JSAC15}.
Except for \cite{Li&Dong:JSAC15}, which considers the storage management design within a finite time period, all the above works consider the long-term average system cost.

Load (demand) scheduling  through demand side management has been studied  by many for shaping the aggregate load at utility through direct load control \cite{ChenWangKishore:TPS14,He&Zhang_TSG12,Koutsopoulos&Tassiulas:JSAC2012} or pricing optimization \cite{Wong&Sen&Ha&Chiang:JSAC2012,SamadiWong:SmartGridComm10}, and at consumer through load scheduling to reduce  electricity bill in response to the dynamic  price \cite{Mohsenian:TSG10,DuLu:TSG11,Kishore:SGC10,KimPoor:TSG11,ChangScaglione:TSG13,YiDong:TSG13}.
With the electricity price known ahead of time, linear programming \cite{Mohsenian:TSG10,DuLu:TSG11 } and DP \cite{Kishore:SGC10} techniques are applied for load scheduling.
Without assuming known future prices, MDP formulation has been considered in \cite{KimPoor:TSG11,ChangScaglione:TSG13}, and opportunistic load scheduling based on optimal stopping rule has been proposed in \cite{YiDong:TSG13}. Combining both utility side and demand side management
is also considered in \cite{Mohsenian&Wang:TSG10,KimRen:JSAC13}, where game theoretic approach is applied for distributed energy management.

Few existing works consider joint optimization of energy storage management and load scheduling. A joint design has been developed in \cite{Chen&Shroff&Ha&Sinha:JSAC2013}, in which the electricity price is assumed to be known ahead of time and the storage model is simplified without the battery operational cost.
Real-time energy storage management with flexible loads has been considered in  \cite{Huang&Mao&Nelms:J_SmartGrid2014} and \cite{SunDongLiang:TSG15}. While \cite{Huang&Mao&Nelms:J_SmartGrid2014} focuses on local demand side, \cite{SunDongLiang:TSG15} combines both grid operator and demand side management using distributed storage. In these works, flexible loads are only modeled in terms of the aggregated energy request; There is no individual task modeling or scheduling being conducted. Furthermore, in these works,  renewable generation, loads, and electricity price are assumed to be independent and identically distributed.
With the increasing penetration of renewable generation and energy storage in the grid, future energy demand and supply are expected to be quite dynamic. Renewable generation, loads, and electricity price may all fluctuate randomly\footnote{Energy pricing design is considered to respond to the energy demand and supply status and to shape the load for demand management. As a result, the real-time price in the future grid could fluctuate much more randomly and quickly, and likely to have complicated statistical behaviors.} with their statistics likely being non-stationary, making them difficult to predict accurately. However, most of existing works design solutions assuming either the future values or statistical knowledge of them to be known.
In addition, although long-term time averaged cost is  typically considered in these existing works, the consumers may prefer a cost saving solution in a period of time defined by their own needs. It is important to provide a solution to meet such need. We aim at proposing a real-time algorithm for joint energy storage and load scheduling to address these issues.

In this paper, we consider joint energy storage management and load scheduling at a residential site equipped with a renewable generator and a storage battery. For renewable source, loads, and electricity price, we assume their dynamics to be arbitrary which  can be non-stationary  and their statistics are unknown. For the residential load, we characterize each individual load task with its own  requested power intensity and service duration, and consider both per load  maximum delay and average delay requirements. For battery storage, we actively model the battery operational constraints and cost due to charging and discharging activities.

We aim at designing a real-time solution for joint energy storage management and load scheduling to minimize the overall system cost over a finite time period, subject to battery operation and load delay constraints. The interaction of load scheduling and energy storage, the finite battery capacity, and finite time period for optimization  complicate the scheduling and energy control decision making over time. To tackle this difficult stochastic problem, we develop techniques through a sequence of problem modification and transformation which enable us to employ Lyapunov optimization to design a real-time  algorithm that otherwise is not directly applicable.
 Interestingly, we show that  the joint load scheduling and energy storage control can be separated and sequentially determined in our real-time optimization algorithm.
Furthermore, both load scheduling and energy control decisions have closed-form solutions, making the real-time algorithm simple to implement. We further show that our proposed real-time algorithm not only provides a bounded performance guarantee to the optimal $T$-slot look-ahead solution which has full future information available, but is also asymptotically equivalent to it as the battery capacity and the considered time period for design  go to infinity.
Simulation results demonstrate the effectiveness of joint load scheduling and energy storage control by our proposed algorithm as compared with
alternative solutions considering neither storage nor scheduling, or storage only.

Different from our recent work \cite{Li&Dong:JSAC15}, in which the energy storage  problem   without flexible loads has been considered, in this work, we explore both energy storage and flexible load scheduling to reduce system cost. Given the individual load modeling, load delay requirements imposed, and the load interaction with energy usage over time, it is highly non-trivial to formulate the joint design problem, develop techniques for a real-time solution,  and provide performance analysis. Through our developed techniques, we show that the joint optimization of load scheduling and storage control can in fact be separated and sequentially solved. Thus, we are able to obtain the load scheduling solution  in closed-form and apply the result in \cite{Li&Dong:JSAC15} for the storage control. Furthermore, we demonstrate that  a lower system cost can be achieved with joint load scheduling and energy storage control than with just energy storage alone.

Comparing with existing works, our proposed algorithm has the following features and advantages: 1) The battery storage operation and associated cost, as well as individual load   and its quality of service, are thoroughly modeled; 2) The algorithm provides a real-time joint solution
for both energy storage control and load task scheduling;
3) The solution
  only relies on the current price, renewable generation, or loads, and does not require any statistical knowledge of them; 4) The solution is designed for a specified period of time which may be useful for practical needs; 5) The solution is provided in closed-form requiring minimum complexity for practical implementation.

The rest of this paper is organized as follows. In Section~\ref{sec:model}, we describe the system model. In Section~\ref{sec:FHA}, we formulate the joint energy management and load scheduling problem. In Section~\ref{sec:RT alg}, we propose a real-time algorithm for our joint optimization problem. In Section~\ref{subsec:PA}, we analyze the performance of algorithm. After presenting our simulation studies in Section~\ref{sec:sim}, we conclude our paper in Section~\ref{sec:conclusion}.

\emph{Notations}:
The main symbols used in this paper are summarized in Table \ref{tab:num}.
\begin{table}[t]
\renewcommand{\arraystretch}{1.4}
\centering
\caption{List of main symbols}\label{tab:num}
\begin{tabular}{|p{0.7cm}|p{7.1cm}|}
\hline
$W_t$ & user's load arriving at time slot $t$ (kWh) \\
\hline
$\rho_{t}$ & load intensity (kWh)\\
\hline
$\lambda_{t}$ & load duration (number of slots)\\
\hline
$d_t$ & delay incurred for $W_t$ before it is served (number of slots) \\
\hline
$d_{t}^{\max}$ & maximum delay  allowed for $W_{t}$ before it is served (number of slots)\\
\hline
$\overline{d_w}$ & average  delay of all arrived loads within the $T_o$-slot period  (number of slots) \\
\hline
$d^{\max}$ & maximum average delay (number of slots) for the loads within the $T_o$-slot period\\
\hline
$C_d(\cdot)$ & cost function associated with the average delay $\overline{d_w}$\\
\hline
$E_t$ & energy purchased from conventional grid at time slot $t$  (kWh) \\
\hline
$E_{\max}$ &  maximum amount of energy that can be bought from the grid per slot~(kWh) \\
\hline
$P_t$ & unit price of buying energy at time slot $t$ ($\$$/kWh) \\
\hline
$P_{\max}$ & maximum unit energy price ($\$$/kWh)\\
\hline
$P_{\min}$ & minimum unit energy price ($\$$/kWh)\\
\hline
$S_t$ & renewable energy harvested at time slot $t$  (kWh) \\
\hline
$S_{w,t}$ & amount of renewable energy directly supplied user's loads to be served at time slot $t$  (kWh)\\
\hline
$S_{r,t}$ & amount  of renewable energy stored into battery at time slot $t$ (kWh)\\
\hline
$Q_t$ & portion of $E_t$ stored into battery at time slot $t$  (kWh)\\
\hline
$R_{\max}$ & maximum charging amount (kWh) per slot allowed for the battery\\
\hline
$D_t$ & amount of energy discharged from the battery at time slot $t$ (kWh) \\
\hline
$D_{\max}$ & maximum discharging amount per slot allowed from the battery (kWh) \\
\hline
$B_t$ & battery energy level at time slot $t$ (kWh) \\
\hline
$B_{\min}$ & minimum energy level required in the battery  (kWh)\\
\hline
$B_{\max}$ & maximum energy level allowed in the battery  (kWh) \\
\hline
$C_{\textrm{rc}}$ & entry cost  for battery due to each charging activity ($\$$)\\
\hline
$C_{\textrm{dc}}$ & entry cost for battery due to each discharging activity ($\$$) \\
\hline
$x_{e,t}$ & entry cost for battery at time slot $t$ as $x_{e,t}= 1_{R,t}C_{\textrm{rc}}+1_{D,t}C_{\textrm{dc}}$ ($\$$) \\
\hline
$x_{u,t}$ & net amount of  energy change in battery at time slot $t$, as $x_{u,t}=
\left|Q_t+S_{r,t}-D_t\right|$ (kWh) \\
\hline
$\overline{x_e}$ & average entry cost  for battery  over the $T_o$-slot period ($\$$)\\
\hline
$\overline{x_u}$ & average net amount of energy change in battery  over the $T_o$-slot period  (kWh) \\
\hline
$C_u(\cdot)$ & cost  function associated with average usage amount $\overline{x_u}$ ($\$$)\\
\hline
$\alpha$ & weight for the cost of scheduling delay\\
\hline
$\abf_t$ & energy storage control action vector at time slot $t$\\
\hline
$\Delta_u$ & desired change  of battery energy level within $T_o$ slots (kWh) \\
\hline
$\mu$ & weight for delay related queues in  Lyapunov function\\
\hline
\end{tabular}
\end{table}

\section{System Model} \label{sec:model}
\begin{figure}
\centerline{
\begin{psfrags}
\psfrag{C}{\normalsize ${P_t}$}
\psfrag{P}{\normalsize ${E_t}$}
\psfrag{S}{\normalsize ${S_t}$}
\psfrag{PQ}{\normalsize ${E_t-Q_t}$}
\psfrag{Q}{\normalsize ${Q_t}$}
\psfrag{S1}{\normalsize ${S_{w,t}}$}
\psfrag{S2}{\normalsize ${S_{r,t}}$}
\psfrag{F}{\normalsize ${D_t}$}
\psfrag{W}{\small \hspace*{-1.5em}${\displaystyle \sum_{\tau=0}^{t}\rho_{\tau}1_{S,t}(d_\tau)}$}
\psfrag{X}{\normalsize $+$}
\psfrag{O}{\normalsize $-$}
\psfrag{Grid}{\small \hspace*{-.5em} Grid}
\psfrag{Loads}{\small \hspace*{-.5em} Loads}
\psfrag{Contrl}{\small \hspace*{-.65em} Contrl.}
\psfrag{Renew.}{\small \hspace*{-.55em} Renew.}
\psfrag{Eng.}{\small \hspace*{-.55em} Gen.}
\psfrag{Battery}{\small \hspace*{-.6em} Battery}
\includegraphics[scale=0.45]{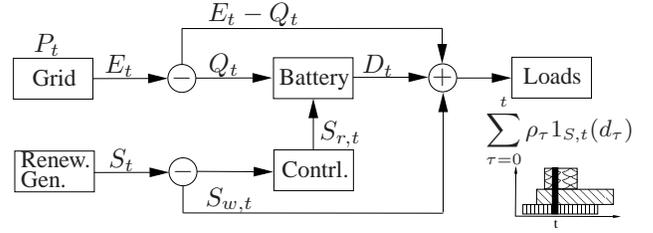}
\end{psfrags}
}
\caption{The residential energy storage  management system.}
\label{fig:single_model}
\end{figure}

We consider a residential-side electricity consuming entity  powered by the conventional grid and a local renewable generator (RG) (\eg wind or solar generators). An energy storage battery is co-located with RG to store energy from both power sources and supply power to the user.
The energy storage  management (ESM) system  is shown in Fig.~\ref{fig:single_model}. As a part of the ESM system, a load scheduling mechanism is implemented to schedule each load task to meet its delay requirements.
We assume the ESM system operates in discrete time slots with $t\in \{0,1,\cdots\}$, and all operations are performed per time slot $t$.
Each component of the EMS system is described below.

\subsection{Load Scheduling}
\begin{figure}[t]
\centerline{
\begin{psfrags}
\psfrag{Wt1}{\normalsize ${W_{t_1}}$}
\psfrag{Wt2}{\normalsize ${W_{t_2}}$}
\psfrag{t1}{\normalsize ${t_1}$}
\psfrag{t2}{\normalsize ${t_2}$}
\psfrag{dt1}{\normalsize ${d_{t_1}}$}
\psfrag{dt2}{\normalsize ${d_{t_2}}$}
\psfrag{rho1}{\normalsize ${\rho_{t_1}}$}
\psfrag{rho2}{\normalsize ${\rho_{t_2}}$}
\psfrag{lmd1}{\normalsize ${\lambda_{t_1}}$}
\psfrag{lmd2}{\normalsize ${\lambda_{t_2}}$}
\psfrag{d}{\normalsize $\ldots$}
\psfrag{dl1}{\normalsize ${d_{t_1}^{\max}}$}
\psfrag{dl2}{\normalsize ${d_{t_2}^{\max}}$}
\psfrag{0}{\small ${0}$}
\psfrag{To}{\small ${T_o-1}$}
\includegraphics[scale=0.6]{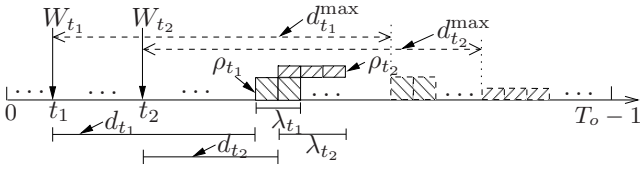}
\end{psfrags}
}
\caption{An example of load scheduling for two arrival loads $W_{t_1}$ and $W_{t_2}$.}
\label{fig:time slot}
\end{figure}

We assume the user has load tasks in various types arriving over time slots. An example of the scheduling time line of two loads is shown in Fig.~\ref{fig:time slot}.
Let $W_t$ denote the load arriving at the beginning of time slot $t$. It is given by $W_{t}=\rho_{t}\lambda_{t}$, where $\rho_t$  and $\lambda_t$ are the load intensity and duration for $W_t$, respectively. We assume $\lambda_t$ is an integer multiple of time slots, and  the minimum duration for any load is $1$, \ie $\lambda_t \in \{1,2,\ldots\}$.
Let $d_{t}^{\max}$ denote the maximum  delay allowed for $W_{t}$ before it is served (multiple of time slots), and let $d_{t}$ denote the actual delay incurred for $W_t$ before it is served. We have
\begin{align}\label{eqn:d_w per time}
d_{t} \in
\{0,1\ldots,d_{t}^{\max}\},\ \forall t.
\end{align}
Thus, the earliest serving time duration for $W_t$ is $[t, t+\lambda_{t}]$, and the latest serving time duration is $[t+d_{t}^{\max}, t+d_{t}^{\max}+\lambda_{t}]$.
We define an indicator function $1_{S,t}(d_\tau)\triangleq\{1: \textrm{if}\ t\in[\tau+d_{\tau},\tau+d_{\tau}+\lambda_{\tau}); \ 0: \textrm{otherwise}\}$, for $\forall\tau\leq t$. It indicates whether or not the load $W_{\tau}$ is being served at time slot $t$.
Consider a $T_o$-slot period. We define $\overline{d_w}$ as the average delay of all arrived loads within this $T_o$-slot period, given by\footnote{Without loss of generality, we start the $T_o$-period at time slot $t=0$.}
\begin{align}\label{eqn:def avg delay}
\overline{d_w}\triangleq\frac{1}{T_o}\sum_{\tau=0}^{T_o-1}d_{\tau}.
\end{align}
Besides the per load maximum delay $d_t^{\max}$ constraint in \eqref{eqn:d_w per time}, we impose a constraint on the average delay $\overline{d_w}$ as
\begin{align}\label{eqn:avg_x_dt}
\overline{d_w}\in\left[0,d^{\max}\right]
\end{align}
where $d^{\max}$ is the maximum average delay for the loads within the $T_o$-slot period. It is straightforward to see that for constraint \eqref{eqn:avg_x_dt} to be effective, we have  $d^{\max}\leq\max_{t\in[0,T_o-1]}\{d_t^{\max}\}$, for $\forall t$.
The average delay $\overline{d_w}$ reflects  the average quality of service for the loads within the $T_o$-slot period.
We define a cost function $C_d(\overline{d_w})$ associated with  $\overline{d_w}$. A longer delay reduces the quality of service and incurs a higher cost. Thus, we assume $C_d(\cdot)$ to be a continuous, convex, non-decreasing function with  derivative $C_d'(\cdot)<\infty$.

\subsection{Energy Sources and Storage}

\subsubsection{Power Sources}
The user can purchase energy from the \emph{conventional grid} with a real-time price. Let $E_t$ denote the amount of energy bought in time slot $t$. It is bounded by
\begin{equation}
\label{equ:Pt_constraint}
E_t\in \left[0,E_{\max}\right]
\end{equation}
where $E_{\max}$ is the maximum amount of energy that can be bought from the grid per slot.。
This amount can be used to directly supply the user's loads and/or be stored into the battery. Let $P_t$ denote the unit price of buying energy at time slot $t$. It is bounded as $P_t\in[P_{\min},P_{\max}]$, where $P_{\min}$ and $P_{\max}$ are the minimum and maximum unit energy prices, respectively. We assume  $P_t$ is known to the user at time slot $t$ and is kept unchanged within the slot duration.
The average cost for the purchased energy from the grid over a $T_o$-slot period is defined by $\overline{J}\defeq \frac{1}{T_o}\sum_{t=0}^{T_o-1}E_tP_t$.

\emph{Renewable generator}:
An RG is used as an alternative energy source in the ESM system.
Let $S_t$ denote the amount of renewable energy harvested at time slot $t$. We assume $S_t$ is first used to supply  the loads scheduled to be served at time slot $t$. Denote this portion by $S_{w,t}$, we have
\begin{align}\label{eqn:S_w bounds}
S_{w,t}=\min\left\{\sum_{\tau=0}^{t}\rho_{\tau}1_{S,t}(d_\tau),S_t\right\}
\end{align}
where the first term in \eqref{eqn:S_w bounds} represents the total energy over those scheduled loads that need to be served at time slot  $t$.
The remaining portion of $S_t$, if any, can be stored into the battery. Since there is a cost associated to the battery charging activity, we use a controller to determine whether or not to store the remaining portion  into the battery.
Let $S_{r,t}$ denote the amount of renewable energy charged into the battery  at time slot $t$. It is bounded by
\begin{align}
\label{eqn:S2-bounds}
S_{r,t} \in \left[0, S_t-S_{w,t}\right].
\end{align}
\subsubsection{Battery Operation}
The battery can be charged from either the grid, the renewable generator, or both at the same time. Let $Q_t$ denote the portion of $E_t$ from the grid that is stored into the battery at time slot $t$. The total charging amount at time slot $t$ is bounded by
\begin{align}\label{eqn:rc_bds}
Q_t+S_{r,t}\in \left[0,R_{\max}\right]
\end{align}
where $R_{\max}$ is the maximum charging amount per slot for the battery.
Similarly, let $D_t$ denote the discharging amount  from the battery at time slot $t$, bounded by
\begin{align}\label{eqn:dc_bds}
D_t \in \left[0, D_{\max}\right]
\end{align}
where $D_{\max}$ is the maximum discharging amount per slot allowed from the battery.
We assume there is no simultaneous charging and discharging activities at the battery, \ie
\begin{align}\label{eqn:rc_dc_constr}
\left(Q_t+S_{r,t}\right)\cdot D_t = 0.
\end{align}
 Let $B_t$ denote the battery energy level at time slot $t$. With a finite capacity,  $B_t$ is bounded by
\begin{equation} \label{eqn:Yt bds}
 B_t\in \left[B_{\min}, B_{\max}\right]
\end{equation}
where $B_{\min}$ and $B_{\max}$ are the minimum energy required and maximum energy allowed in the battery, respectively.
The dynamics of $B_t$ over time due to charging and discharging activities are given by\footnote{We consider an ideal battery model with no leakage of stored energy over time and full charging/discharging efficiency.
}
\begin{equation}
\label{eqn:dynamics of SOB}
B_{t+1}=B_t+Q_t+S_{r,t}-D_t.
\end{equation}

It is known from battery technology that frequent charging/discharging activities cause a battery to degrade over time \cite{Ramadrass:JPS02,MaZou:CST15,Michelusi:COM13}.\footnote{The lifetime of battery can be coarsely measured by the number of full charge cycles.}
Both the frequency of charging or discharging and the amount that is charged or discharged affect the battery lifetime. Given this, we model two types of battery operational costs associated with the charging/discharging activities: \emph{entry cost} and \emph{usage cost}.

The entry cost is a fixed cost incurred due to each charging or discharging activity.
Define two indicator functions to represent charging and discharging activities as ${1_{R,t}\triangleq\{1: \textrm{if}\ Q_t+S_{r,t}>0; \; 0: \textrm{otherwise}\}}$ and ${1_{D,t}\triangleq\{1: \textrm{if}\ D_t>0; \; 0: \textrm{otherwise}\}}$, respectively.
Let $C_{\textrm{rc}}$ denote the entry cost for each charging activity and $C_{\textrm{dc}}$ for that of the discharging activity.
Let $x_{e,t}$ denote the entry cost at time slot $t$. It is given by $x_{e,t}\triangleq 1_{R,t}C_{\textrm{rc}}+1_{D,t}C_{\textrm{dc}}$.
We define the time-averaged entry cost over the $T_o$-slot period as $\overline{x_e} \triangleq \frac{1}{T_o}\sum_{t=0}^{T_o-1}x_{e,t}$.

The usage cost is defined as the cost associated with the battery charging and discharging amount.
Let $x_{u,t}\defeq|Q_t+S_{r,t}-D_t|$ denote the net amount of energy change in battery  at time slot $t$ due to charging or discharging.
From \eqref{eqn:rc_bds} and \eqref{eqn:dc_bds}, it follows that $x_{u,t}$ is bounded by
\begin{align} \label{eqn:x_u}
x_{u,t} \in \left[0,\max\left\{R_{\max},D_{\max}\right\}\right].
\end{align}
In general, the battery usage cost is associated with charge cycles\footnote{Charging the battery and then discharging it to the same level is considered a charge cycle.}.
Each charge cycle typically lasts for a period of time in a day \cite{Duggal:PWRS15}.
To approximate this, we consider the average net amount of energy change in the battery over the $T_o$-slot period, defined as $\overline{x_u}\triangleq\frac{1}{T_o}\sum_{t=0}^{T_o-1}x_{u,t}$.
From \eqref{eqn:x_u}, it is straightforward to see that $\overline{x_u}$ is bounded by $\overline{x_u}\in \left[0,\max\left\{R_{\max},D_{\max}\right\}\right]$.
We model the usage cost  as a function of $\overline{x_u}$, denoted by $C_u(\overline{x_u})$.
It is known that faster charging/discharging within a fix period
has a more detrimental effect on the life time of the battery. Thus, we assume $C_u(\overline{x_u})$ is a continuous, convex, non-decreasing function with  derivative $C_u'(\overline{x_u})<\infty$.\footnote{Such a convex cost function has also been adopted in literature \cite{GuoFang:TSG13,MaZou:TCST13}.}

Based on the above, the average battery operational cost over the $T_o$-slot period due to charging/discharging activities is given by $\overline{x_e}+C_u(\overline{x_u})$.

\subsection{Supply and Demand Balance}
For each load $W_{\tau}$ arrived at time slot $\tau$, if it is scheduled to be served at time slot $t$ ($\geq \tau$), the energy supply needs to meet the amount  $\rho_{\tau}$ scheduled for $W_{\tau}$.
The overall energy supply must be equal to the total demands from those loads which need to be served at time slot $t$.
Thus, we have supply and demand balance relation given by
\begin{align}\label{eqn:Wt_constraint}
E_t-Q_t+S_{w,t}+D_t=
\sum_{\tau=0}^{t}\rho_{\tau}1_{S,t}(d_\tau),&\quad \forall t.
\end{align}

\section{Joint Energy Storage Management and Load Scheduling:\ Problem Formulation}\label{sec:FHA}
Our goal is to jointly optimize the load scheduling and energy flows and storage control for the ESM system to minimize an overall system cost over the $T_o$-slot period.
The loads, renewable generation, and price $\{W_t,S_t,P_t\}$ have complicated statistical behaviors which may be non-stationary and thus are often difficult to acquire or predict in practice.  In our design, we assume arbitrary dynamics for $\{W_t,S_t,P_t\}$ and do not assume their statistical knowledge being known. We intend to develop a real-time control algorithm that is capable to handle such arbitrary and unknown system inputs.

We model the overall system cost as a weighted sum of the  cost from energy purchase and battery degradation, and the cost of scheduling delay.
Define $\abf_t\triangleq[E_t,Q_t,D_t,S_{w,t},S_{r,t}]$ as the control action vector for the energy flow in the ESM system at time slot $t$.
Our goal is to find an optimal policy $\{\abf_t,d_t\}$ that minimizes the time-averaged system cost.
This optimization problem is formulated as follows
\begin{align}
\textrm{\bf P1:}\; \min_{\{\abf_t,d_t\}} &\;\;
\overline{J}+\overline{x_e}+ C_u(\overline{x_u})+\alpha C_d(\overline{d_w})\nn\\
\textrm{s.t.} \;\;&
\eqref{eqn:d_w per time},\eqref{eqn:avg_x_dt},\eqref{equ:Pt_constraint},\eqref{eqn:S2-bounds},\eqref{eqn:rc_dc_constr},
\eqref{eqn:Wt_constraint}, \textnormal{and} \nonumber\\
&0\leq S_{r,t}+Q_t\leq \min\left\{R_{\max},B_{\max}-B_t\right\}\label{eqn:rc_bds_strict}\\
&0\leq D_t \leq \min\left\{D_{\max},B_t-B_{\min}\right\}\label{eqn:dc_bds_strict}
\end{align}
where $\alpha$ is the positive weight for the cost of scheduling delay;
It sets  the relative weight between energy related cost and load delay incurred by scheduling in the joint optimization.

Note that in {\bf P1}, $\{W_t,S_t,P_t\}$ are random, and their future values are unknown at time slot $t$. Thus, {\bf P1} is a finite time horizon joint stochastic optimization problem which is difficult to solve. Joint energy storage control and load scheduling complicates the problem, making it much more challenging than each separate problem alone. The finite battery capacity imposes a hard constraint on the control actions $\{\abf_t\}$, making $\{\abf_t\}$ correlated over time, due to the time-coupling dynamics of $B_t$ in \eqref{eqn:dynamics of SOB}.
Furthermore, the finite time horizon problem is much more difficult to tackle than the infinite time horizon problem as considered in most existing energy storage works. New techniques need to be developed for a real-time control solution.

In the following, we focus on proposing a real-time algorithm to provide a suboptimal solution to {\bf P1} with a certain performance guarantee. To do this, we first modify {\bf P1} to allow us to design a real-time algorithm for joint energy storage control and load scheduling at every time slot. We later discuss how our  solution can meet the constraints of {\bf P1}.

\subsection{Problem Modification}
Due to the finite battery capacity constraint, the control actions $\{\abf_t\}$ are coupled over time.
To remove the time coupling, similar to the technique used in our previous work   \cite{Li&Dong:JSAC15} for energy storage only problem, we remove the finite battery capacity constraint, and instead we impose a constraint on the change of battery energy level over the $T_o$-slot period.\ Specifically, by \eqref{eqn:dynamics of SOB}, the change of battery energy level over the $T_o$-slot period is $B_{T_o}-B_0=\sum_{t=0}^{T_o-1}\left(Q_{t}+S_{r,t}-D_t\right)$.
We now set this change to be a desired value $\Delta_u$, \ie
\begin{align}
\label{eqn:delta_u}
\frac{1}{T_o}\sum_{t=0}^{T_o-1}\left(Q_t+S_{r,t}-D_t\right) = \frac{\Delta_u}{T_o}.
\end{align}
Note that, $\Delta_u$ is only a desired value we set, which may not be achieved by a control algorithm at the end of $T_o$-slot period.
We will quantify the amount of mismatch with respect to $\Delta_u$ under our proposed control algorithm in Section~\ref{subsec:PA}. By the battery capacity and (dis)charging constraints, it is easy to see that $|\Delta_u|\leq\Delta_{\max}\defeq \min\{B_{\max}-B_{\min},T_o\max\{R_{\max},D_{\max}\}\}$.

We now modify {\bf P1} to the follow optimization problem by adding  the new constraint \eqref{eqn:delta_u}, and removing the battery capacity constraint
\eqref{eqn:Yt bds}\begin{align}
\textrm{\bf P2:}\; &\min_{\{\abf_t,d_t\}} \;\;
\overline{J}+\overline{x_e}+C_u(\overline{x_u})+\alpha C_d(\overline{d_w})\nn\\
\rm{s.t.} \;\;
&\eqref{eqn:d_w per time},\eqref{eqn:avg_x_dt}-\eqref{eqn:rc_dc_constr},\eqref{eqn:Wt_constraint}, \eqref{eqn:delta_u}. \nonumber
\end{align}
Note that by removing the battery capacity constraint \eqref{eqn:Yt bds}, we remove the dependency of per-slot charging/discharging amount on $B_t$ in constraints \eqref{eqn:rc_bds_strict} and \eqref{eqn:dc_bds_strict}, and replace them by \eqref{eqn:rc_bds} and \eqref{eqn:dc_bds}, respectively.

\subsection{Problem Transformation}
In {\bf P2}, both battery average usage cost $C_u(\overline{x_u})$ and scheduling delay cost $C_d(\overline{d_w})$ are functions of time-averaged variables, which complicates the problem.
Using the technique introduced in \cite{Neely:ArXiv2010}, we now transform the problem into  one that only contains the time-average of the functions.
Specifically, we introduce auxiliary variables ${\gamma_{u,t}}$ and $\gamma_{d,t}$ for $x_{u,t}$ and $d_t$, respectively, and impose the following constraints
\begin{align}
&0\le \gamma_{u,t} \le \max\{R_{\max},D_{\max}\}, \ \forall t\label{eqn:gamma_bds}\\
&\overline{\gamma_u}=\overline{x_u} \label{avg_r=avg_x}\\
&0\leq\gamma_{d,t}\leq \min\{d_{t}^{\max},d^{\max}\}, \ \forall t\label{avg_gamma2b}\\
&\overline{\gamma_d}=\overline{d_w}\label{eqn:gamma2_bds}
\end{align}
where $\overline{\gamma_i}\defeq \frac{1}{T_o}\sum_{\tau=0}^{T_o-1}\gamma_{i,t}$, for $i=u,d$.
The above constraints ensure that each auxiliary variable lies  in the same range as its original variable, and its time average  is the same as that of its original variable.
Define $\overline{C_i(\gamma_i)}\triangleq \frac{1}{T_o}\sum_{t=0}^{T_o-1}C_i(\gamma_{i,t})$ as the time average of $C_i(\gamma_{i,t})$ over $T_o$ slots, for $i=u,d$.
Applying \eqref{avg_r=avg_x} and \eqref{eqn:gamma2_bds} to the objective of {\bf P2}, and defining  $\pibf_t\triangleq[\abf_t,d_t,\gamma_{u,t},\gamma_{d,t}],$ we transform  {\bf P2} into the following optimization problem
\begin{align}
\textrm{\bf P3:}\; &\min_{\{\pibf_t\}} \;\;
\overline{J}+\overline{x_e}+ \overline{C_u(\gamma_u)}+\alpha\overline{C_d(\gamma_d)}\nn\\
\rm{s.t.} \;\;
&\eqref{eqn:d_w per time},\eqref{eqn:avg_x_dt}-\eqref{eqn:rc_dc_constr},
\eqref{eqn:Wt_constraint}, \eqref{eqn:delta_u}-\eqref{eqn:gamma2_bds}\nonumber
\end{align}
where the terms in the objective are all $T_o$-slot time-averaged cost functions. We can show that  {\bf P2} and {\bf P3} are equivalent, \ie they have the same optimal control solution $\{\abf_t^*,d_t^*\}$ (See Appendix~\ref{appA}).

Although {\bf P3} is still difficult to solve, it enables us to design a dynamic control and scheduling algorithm for joint energy storage control and load scheduling by adopting Lyapunov optimization technique \cite{book:Neely}. In the following, we propose our real-time algorithm for {\bf P3}, and then design parameters to ensure the proposed solution meets the battery capacity constraint in the original  {\bf P1} which is removed in {\bf P2}.

\section{Joint Energy Storage Management and Load Scheduling: Real-Time Algorithm  }\label{sec:RT alg}
By Lyapunov optimization, we first introduce virtual queues for each time-averaged inequality and equality constraints of {\bf P3} to transform them into queue stability problems. Then, we design a real-time algorithm based on the drift of Lyapunov function defined on these virtual queues.

\subsection{Virtual Queues}
We introduce a virtual queue $X_t$ to meet constraint \eqref{eqn:avg_x_dt}, evolving as follows
\begin{align}\label{eqn:queue X}
X_{t+1}=\max\left(X_t+d_{t}-d^{\max},0\right).
\end{align}
From \eqref{eqn:def avg delay}, the above results in $\overline{d_w}\leq d^{\max}+(X_{T_o}-X_0)/{T_o}$. Thus, formulating the virtual queue $X_t$  in \eqref{eqn:queue X} will guarantee to meet the average delay constraint \eqref{eqn:avg_x_dt} with a margin $(X_{T_o}-X_0)/{T_o}$.
In Section~\ref{subsec:PA}, we will further discuss this constraint under our proposed algorithm.

For constraint \eqref{eqn:delta_u}, dividing both sides by $T_o$ gives the time-averaged net change of battery energy level per slot being $\Delta_u/T_o$. To meet this constraint, we introduce a virtual queue $Z_t$, evolving as follows
\begin{align}\label{eqn:queue Z}
Z_{t+1}=Z_t+Q_t+S_{r,t}-D_t-\frac{\Delta_u}{T_o}.
\end{align}
From $B_t$ in \eqref{eqn:dynamics of SOB} and $Z_t$ above, we can show that they are different by a time-dependent shift as follows \begin{align} \label{eqn:A_t}
Z_t=B_t-A_t, \textrm{ where } A_t\defeq A_o+\frac{\Delta_u}{T_o}t.
\end{align}
The linear time function $\frac{\Delta_u}{T_o}t$ in $A_t$ is to ensure that  constraint \eqref{eqn:delta_u} is satisfied.
Due to this shift $A_t$, the range of $Z_t$ is expanded to the entire real line, \ie $Z_t\in \mathbb{R}$ for $B_t \in \mathbb{R}^+$. Note that $A_o$ is a design parameter. Later, we design $A_o$ to ensure that our control solution $\{\abf_t\}$ for the energy flows in our proposed algorithm satisfies the battery capacity constraint \eqref{eqn:Yt bds} imposed in {\bf P1}.

Finally, to meet  constraints \eqref{avg_r=avg_x} and \eqref{avg_gamma2b}, we establish virtual queues $H_{u,t}$ and $H_{d,t}$, respectively, as follows
\begin{align}
H_{u,t+1}&=H_{u,t}+\gamma_{u,t}-x_{u,t} \label{eqn:queue H1}\\
H_{d,t+1}&=H_{d,t}+\gamma_{d,t}-d_{t}. \label{eqn:queue H2}
\end{align}

From Lyapunov optimization, it can be shown that
satisfying constraints \eqref{eqn:avg_x_dt}, \eqref{eqn:delta_u}, \eqref{avg_r=avg_x}, and \eqref{avg_gamma2b} is equivalent to maintaining
the stability of queues $X_t$, $Z_t$, $H_{u,t}$, and $H_{d,t}$, respectively\cite{book:Neely}.

\subsection{Real-Time Algorithm} \label{subsec:algorithm}

Note that $Z_t$ and $H_{u,t}$ are the virtual queues related to the battery operation, while $X_t$ and $H_{d,t}$ are  those related to the scheduling delay. Let $\Thetabf_t\triangleq[Z_t,H_{u,t},X_t,H_{d,t}]$ denote the virtual queue vector.
We define the quadratic Lyapunov function $L(\Thetabf_t)$ for $\Thetabf_t$ as follows
\begin{align}\label{eqn:lyapunov function}
L(\Thetabf_t)\triangleq\frac{1}{2}\left[Z_t^2+ H_{u,t}^2+\mu\left(X_t^2+H_{d,t}^2\right)\right]
\end{align}
where $\mu$ is a positive weight  to adjust the relative importance of load delay related queues in the Lyapunov function.
We define a one-slot sample path Lyapunov drift
as $\Delta(\Thetabf_t)\triangleq L\left(\Thetabf_{t+1}\right)-L(\Thetabf_{t})$,
which only depends on the current system inputs $\{W_t,S_t,P_t\}$.

Instead of directly minimizing the system cost objective in {\bf P3}, we consider the  \emph{drift-plus-cost} metric given by $\Delta(\Thetabf_t)+V[E_tP_t+x_{e,t}\allowbreak+C_u(\gamma_{u,t})\allowbreak+\alpha C_d(\gamma_{d,t})]$. It is a weighted sum of the  drift $\Delta(\Thetabf_t)$ and the system cost at time slot $t$ with $V>0$ being the relative weight between the two terms.

Directly using the drift-plus-cost function to determine control action $\pibf_t$ is still challenging. In the following, we use an upper bound of this drift-plus-cost function to design our real-time algorithm.
The upper bound is derived in Appendix \ref{appB} as \eqref{eqn:lemma drift bound}. Using this upper bound, we formulate a per-slot real-time optimization problem and solve it at every time slot $t$. By removing all the constant terms independent of control action $\pibf_t$, we arrive at the following optimization problem
\begin{align}
\textrm{\bf P4}: \; \min_{\pibf_t} \;
 &Z_t\left[E_t+S_{r,t}+S_{w,t}-\rho_{t}1_{S,t}(d_t)\right]- \left|H_{u,t}\right|S_{w,t}\nn\\
 &\hspace*{-0.5em}+H_{u,t}\left[\gamma_{u,t}-(E_t+S_{r,t})\right]+\left|H_{u,t}\right|\rho_{t}1_{S,t}(d_t)\nn\\
 &\hspace*{-0.5em}+\mu X_td_{t}+\mu H_{d,t}(\gamma_{d,t}-d_{t})\nn\\
 &\hspace*{-0.5em}+V\left[E_tP_t+x_{e,t}+ C_{u}(\gamma_{u,t})+\alpha C_{d}(\gamma_{d,t})\right] \nn\\
\rm{s.t.} &\;\;
\eqref{eqn:d_w per time},\eqref{equ:Pt_constraint}-\eqref{eqn:rc_dc_constr},
\eqref{eqn:Wt_constraint}, \eqref{eqn:gamma_bds},\eqref{avg_gamma2b}.\nn
\end{align}
Note that the term  $\sum_{\tau=0}^{t}\rho_{\tau}1_{S,t}(d_\tau)$ in the upper bound \eqref{eqn:lemma drift bound} is the total energy demand from the scheduled loads at time slot $t$.  Since delay $d_\tau$ for $\tau\in\{0,1,\ldots,t-1\}$ are determined in previous time slot $\tau \le t-1$ by solving {\bf P4},  only $\rho_t1_{S,t}(d_t)$ is a function of $\pibf_t$ at time slot $t$, and is part of the objective of {\bf P4}.

Denote the optimal solution of {\bf P4} by $\pibf_t^*\triangleq[\abf_t^*,d_t^*,\gamma_{u,t}^*,\gamma_{d,t}^*]$.
After regrouping the terms in the objective of {\bf P4} with respect to different control variables, we show that {\bf P4} can be separated into four  sub-problems to be solved  sequentially and variables in $\pibf_t^*$ can be determined separately. The steps are described below.
\begin{list}{\labelitemi}
{
\setlength\leftmargin{2em}
\setlength\labelwidth{1.5em}
\setlength\labelsep{.5em}
}
\item[\it S1)]Determine $d_t^*$ and $\gamma_{d,t}^*$ by solving the following ${\bf \text{\bf P4}_{a1}}$ and ${\bf \text{\bf P4}_{a2}}$, respectively.
 \begin{align}
   {\bf \text{\bf P4}_{a1}:}  &\min_{d_t}
  \ \mu d_{t}(X_t-H_{d,t})-\rho_{t}1_{S,t}(d_t)\left(Z_t-|H_{u,t}|\right) \nn\\
   &\textrm{s.t.} \ \eqref{eqn:d_w per time}. \nn \\
 {\bf \text{\bf P4}_{a2}:}\; &\min_{\gamma_{d,t}}
  \ \mu H_{d,t}\gamma_{d,t}+V\alpha C_d(\gamma_{d,t}) \quad \textrm{s.t.} \;\; \eqref{avg_gamma2b}. \nn
  \end{align}
\item[ \it S2)] Determine $S_{w,t}^*$ in \eqref{eqn:S_w bounds} using $d_t^*$ obtained in S1).
\item[ \it S3)] Using $S_{w,t}^*$ obtained in S2) in  \eqref{eqn:Wt_constraint}, determine $\gamma_{u,t}^*$ and $\abf_t^*$ by solving the following ${\bf \text{\bf P4}_{b1}}$ and ${\bf \text{\bf P4}_{b2}}$, respectively.
  \begin{align}
  {\bf \text{\bf P4}_{b1}:} \; &\min_{\gamma_{u,t}} \;  H_{u,t}\gamma_{u,t}+V C_u(\gamma_{u,t}) \quad \textrm{s.t.} \;\; \eqref{eqn:gamma_bds}. \nn \\
  {\bf \text{\bf P4}_{b2}:}\; &\min_{\abf_t} \;
  E_t(Z_t-H_{u,t}+VP_t)+S_{r,t}(Z_t-H_{u,t}) \nn\\
  &\hspace*{1em}\quad +V\left(1_{R,t}C_{\textrm{rc}}+1_{D,t}C_{\textrm{dc}}\right) \nn\\
  &\textrm{s.t.} \quad
  \eqref{equ:Pt_constraint}-\eqref{eqn:rc_dc_constr}, \eqref{eqn:Wt_constraint}.\nn
  \end{align}
\end{list}

\emph{Remark:} An important and interesting observation of the above is that the joint optimization of load scheduling and energy storage control can in fact be separated: The scheduling decision $d_t^*$ is determined first in ${\bf \text{\bf P4}_{a1}}$. Based on the resulting energy demand in the current time slot $t$, energy storage control decision  $\abf_t^*$ is then determined in ${\bf \text{\bf P4}_{b2}}$. Note that the  two sub-problems are interconnected through the current virtual queue backlogs $Z_t$ and $H_{u,t}$ for the battery energy level $B_t$ and battery energy net change $x_{u,t}$, respectively. The storage control decision $\abf_t^*$ will further change the battery energy level and affect  $H_{u,t}$ at the next time slot. Thus, although the scheduling and storage decisions are separately determined, they are interconnected through battery energy level and usage, and sequentially influence each other.

In the following, we solve each subproblem and obtain a closed-form solution. As a result, the optimal solution $\pibf_t^{*}$ is obtained in closed-form.

\subsubsection{The optimal $d_t^*$}

The optimal scheduling delay $d_t^*$ for  ${\bf \text{\bf P4}_{a1}}$ is given below.
\begin{proposition}\label{prop dt}
Let $\omega_o \triangleq -\rho_{t}\left(Z_t-|H_{u,t}|\right)$, $\omega_1\triangleq\mu\left(X_t-H_{d,t}\right)$, and $\omega_{d^{\max}_t}\triangleq\mu d^{\max}_t(X_t-H_{d,t})$.
\begin{enumerate}
\item If $X_t-H_{d,t}\ge 0$, then
$d_t^*=\begin{cases}  0 & \textrm{if } \omega_o \le \omega_1 \\  1 & \textrm{otherwise};
\end{cases}
$
\item If $X_t-H_{d,t} < 0$, then
$d_t^*=\begin{cases}   0 & \textrm{if } \omega_o \le \omega_{d^{\max}_t} \\ \  d_t^{\max} & \textrm{otherwise}.
\end{cases}
$
\end{enumerate}
\end{proposition}
\IEEEproof
See Appendix~\ref{app:prop dt}.

\emph{Remark}: Note that $\omega_o$, $\omega_1$, and $\omega_{d^{\max}_t}$ are the objective values of ${\bf \text{\bf P4}_{a1}}$ when $d_t=0, 1$, and $d^{\max}_t$, respectively. Furthermore, $w_o$ depends on the virtual queue backlogs ($Z_t$ and $H_{u,t}$) related to the battery energy level, while  $\omega_1$ and $\omega_{d^{\max}_t}$ depend on the virtual queue backlogs ($X_t$ and $H_{d,t}$) related to load delay. Proposition~\ref{prop dt} shows that the scheduling decision for load $W_t$ is to either  immediately serve it ($d^*_t=0$) or delay its serving time ($d^*_t=1$ or $d^{\max}_t$).  This decision depends on whether  the battery energy is high enough (so $W_t$ will be served immediately)\ or the delays for the scheduled loads so far are low enough (so $W_t$ will be delayed). When the load is delayed to serve, the delay should be either minimum or maximum depending on the existing scheduling delays of the past loads.

\subsubsection{The optimal $\gamma_{d,t}^*$ and $\gamma_{u,t}^*$}
Since $C_d(\cdot)$ and $C_u(\cdot)$ are both convex, the objectives of ${\bf \text{\bf P4}_{a2}}$ and ${\bf \text{\bf P4}_{b1}}$ are convex.
Let $C_i'(\cdot)$ denote the first derivative of $C_i(\cdot)$, and $C_i'^{-1}(\cdot)$ denote the inverse function of $C_i'(\cdot)$, for $i=d,u$.
We obtain the optimal solutions $\gamma^*_{i,t}$, $i=d,u$, for ${\bf \text{\bf P4}_{a2}}$ and ${\bf \text{\bf P4}_{b1}}$ as follows.
\begin{lemma}\label{lemma33}
The optimal  $\gamma^*_{i,t}$, for $i=d,u$, is given by
\begin{align}
\label{eqn:optimal gamma}
\gamma^*_{i,t}=
    \begin{cases}
        0&\text{if}\ H_{i,t}\geq 0\\
        \Gamma_i& \text{if}\ H_{i,t}< -V\beta_i C_i'(\Gamma_i)\\
        C_i'^{-1}\left(-\frac{H_{i,t}}{V\beta_i}\right)&\text{otherwise}.
    \end{cases}
\end{align}
where $\beta_u=1$, $\beta_d = {\alpha}/{\mu}$, $\Gamma_u\triangleq\max\{R_{\max},D_{\max}\}$, and $\Gamma_d\triangleq\min\{d_t^{\max},d^{\max}\}$.
\end{lemma}
\IEEEproof
See Appendix~\ref{app:prop1}.

\subsubsection{The optimal $\abf^*_t$}\label{subsubsec:a_t*}
Once the scheduling decision $d^*_t$ for $W_t$ is determined, the total energy demand from the scheduled loads, \ie $\sum_{\tau=0}^{t}\rho_{\tau}1_{S,t}(d_\tau^*)$, is determined.
Given this energy demand,  ${\bf \text{\bf P4}_{b2}}$ is solved to obtain the optimal control solution $[E_t^*,Q_t^*,D_t^*,S_{r,t}^*]$ in $\abf^*_t$.
This subproblem for energy storage and control  is essentially the same as in \cite{Li&Dong:JSAC15}, where the energy storage only problem is considered.
Thus, the closed-form solution can be readily obtained from \cite{Li&Dong:JSAC15}. Here we directly state the result.

Define $L_t^*\triangleq \sum_{\tau=0}^{t}\rho_{\tau}1_{S,t}(d_\tau^*)$ as the current energy demand at time slot $t$. Define the idle state of the battery as the state where there is no charging or discharging activity.
The control solution under this idle state is denoted by $[E_t^\textrm{id}, Q_t^\textrm{id},D_t^\textrm{id},S_{r,t}^\textrm{id}]$.
By supply-demand balancing equation \eqref{eqn:Wt_constraint}, it is given by $E_t^\textrm{id}=L^*_t-S_{w,t}^*$, $Q_t^\textrm{id}=D_t^\textrm{id}=S_{r,t}^\textrm{id}=0$.
Let $\xi_t$ denote the  objective value in ${\bf \text{\bf P4}_{b2}}$ for the battery being in the idle state.
We have $\xi_t= (L^*_t-S_{w,t}^*)(Z_t-H_{u,t}+VP_t)$.
Denote $\abf_t'=[E_t',Q_t',D_t',S_{w,t}^*,S_{r,t}']$. The optimal control solution $\abf^*_t$ of $\bf \text{\bf P4}_{b2}$ is given in three cases below.

\newcounter{qcounter}
\begin{list}{{\it \roman{qcounter}$\left.\right)$~}}
{\usecounter{qcounter}
\setlength\leftmargin{1.5em}
\setlength\labelwidth{2em}
\setlength\labelsep{0em}
\setlength\itemsep{.5em}
}
\item {\it For $Z_t-H_{u,t}+VP_t\leq 0$}: The battery is in either charging or  idle state. The solution $\abf_t'$ in charging state is given by
$\begin{cases}
      D'_t=0,\\
      S'_{r,t}=\min\left\{S_t-S_{w,t}^*,R_{\max}\right\} \\
      Q'_t=\min\left\{R_{\max}-S'_{r,t},E_{\max}-L_t^*+S_{w,t}^*\right\} \\
      E'_t=\min\left\{L_t^*+R_{\max}-S_{w,t}^*-S'_{r,t},E_{\max}\right\}
      \end{cases}.
$

If $E'_t(Z_t-H_{u,t}+VP_t)+(Z_t-H_{u,t})S'_{r,t}+VC_{\textrm{rc}}1_{R,t} <\xi_t$,
      then $\abf_t^*=\abf_t'$; Otherwise, $\abf_t^*=\abf_t^\textrm{id}$.
\item {\it For $Z_t-\mu_uH_{u,t}<0\leq Z_t-H_{u,t}+VP_t$}: The battery is either in  charging, discharging, or idle state. The solution $\abf_t'$ in charging or discharging state is given by
$      \begin{cases}
      D'_t=\min\left\{L_t^*-S_{w,t}^*,D_{\max}\right\}\\
      S'_{r,t}=\min\left\{S_t-S_{w,t}^*,R_{\max}\right\}\\
      Q'_t=0,\\
      E'_t=\left[L_t^*-S_{w,t}^*-D_{\max}\right]^+
      \end{cases}.
$

      If $E'_t(Z_t-H_{u,t}+VP_t)+(Z_t-H_{u,t})S'_{r,t}+V(C_{\textrm{rc}}1_{R,t}+C_{\textrm{dc}}1_{D,t})
      <\xi_t$, then $\abf_t^*=\abf_t'$;  Otherwise, $\abf_t^*=\abf_t^\textrm{id}$.
\item {\it For $0\leq Z_t-H_{u,t}<Z_t-H_{u,t}+VP_t$}: The battery is in either discharging or idle state. The solution $\abf_t'$ in discharging state is given by
$      \begin{cases}
      D'_t=\min\left\{L_t^*-S_{w,t}^*,D_{\max}\right\}\\
      S'_{r,t}= Q'_t=0, \\
      E'_t=\left[L_t^*-S_{w,t}^*-D_{\max}\right]^+
      \end{cases}.
$

      If ${E'_t(Z_t-H_{u,t}+VP_t)+VC_{\textrm{dc}}1_{D,t}<\xi_t}$, then  $\abf_t^*=\abf_t'$;
      Otherwise, $\abf_t^*=\abf_t^\textrm{id}$.
\end{list}

In each  case above, the cost of charging or discharging is compared with the cost $\xi_t$ of being in an idle state, and the control solution of  $\bf \text{\bf P4}_{b2}$ is the one with the minimum cost.
The condition for each case depends on $Z_t$, $H_{u,t}$ and $P_t$, where $Z_t$ and $H_{u,t}$ are rated to battery energy level $B_t$ and usage cost $x_{u,t}$, respectively. Thus, Cases i)-iii) represent the control actions at different battery energy levels (\ie low, moderate, or high) and electricity prices.

\subsubsection{Feasibility of $\abf^*_t$ to {\bf P1}}
Recall that we have removed the battery capacity constraint \eqref{eqn:Yt bds} when modifying {\bf P1} to {\bf P2}. Thus, this constraint is no longer imposed in ${\bf \text{\bf P4}_{b2}}$, and our real-time algorithm may not provide feasible control solutions $\{\mathbf{a}^*_t\}$ to {\bf P1}.
To ensure the solution is still feasible to the original problem {\bf P1}, we design our control parameters $A_o$ and $V$.
The result readily follows \cite[Proposition 2]{Li&Dong:JSAC15}. We omit the details and only state the final result below.
\begin{proposition} \label{prop1}
For the optimal solution $\abf_t^*$ of ${\bf \text{\bf P4}_{b2}}$, the resulting $B_t$ satisfies the battery capacity constraint \eqref{eqn:Yt bds}, and $\{\abf_t^*\}$ is feasible to {\bf P1}, if $A_o$ in \eqref{eqn:A_t} is given by
\begin{align}\label{eqn:A_o}
A_o= \begin{cases} A_o' & \textrm{if}\ \Delta_u\geq0 \\ A_o'-\Delta_u & \textrm{if}\ \Delta_u<0 \end{cases}
\end{align}
where $A_o'=B_{\min}+VP_{\max}+VC_u'(\Gamma_u)+\Gamma_u+D_{\max}+\frac{\Delta_u}{T_o}$,
and $V\in [0,V_{\max}]$ with
\begin{align}\label{eqn:V_max}
\hspace*{-1em}V_{\max}=\frac{B_{\max}-B_{\min}-R_{\max}-D_{\max}-2\Gamma_u-|\Delta_u|}{P_{\max}+C_u'(\Gamma_u)}.
\end{align}
\end{proposition}

\subsection{Discussions}
We summarize the proposed real-time joint load scheduling and energy storage control  in Algorithm~\ref{alg1}. Due to the separation of joint optimization, the algorithm provides a clear sequence of control decisions at each time slot $t$.
\begin{algorithm}[t]
\caption{Real-Time Joint Load Scheduling and Energy Storage Management}\label{alg1}
Set the desired value of $\Delta_u$. Set $A_o$ and $V$ as in \eqref{eqn:A_o} and \eqref{eqn:V_max}, respectively.

At time slot $0$: Set $Z_0=X_0=H_{u,0}=H_{d,0}=0$.\\
At time slot $t$: Obtain the current values of $\{W_t, S_t, P_t\}$
 \begin{enumerate}
\item  \emph{Load scheduling:} Determine $d_t^*$ according to Proposition~\ref{prop dt} and $\gamma_{d,t}^*$ according to \eqref{eqn:optimal gamma}, respectively.
\item \emph{Energy Storage Control:}
\begin{enumerate}
\item {Renewable contribution:}
Determine $S_{w,t}^*$ in \eqref{eqn:S_w bounds} using $d_t^*$ obtained above.
\item {Energy purchase and storage:} Determine $\gamma_{u,t}^*$ according to \eqref{eqn:optimal gamma} and $\abf_t^*$ according to Cases i)-iii) in Section~\ref{subsubsec:a_t*}.
\end{enumerate}
\item \emph{Updating virtual queues:} Use $\pibf_t^*$ to update $B_t$ based on \eqref{eqn:dynamics of SOB}, and $X_t,Z_t,H_{u,t},H_{d,t}$ based on \eqref{eqn:queue X}$-$\eqref{eqn:queue H2}.
\end{enumerate}
\end{algorithm}

Recall that we modify the original joint optimization problem {\bf P1} to {\bf P3}, and apply Lyapunov optimization to propose a real-time algorithm for {\bf P3} which is to solve the per-slot optimization problem {\bf P4}. We then design system parameters $A_o$ and
$V_{\max}$ to ensure our solution satisfies the battery capacity constraint, which is removed when we modify {\bf P1} to {\bf P4}. As a result, our proposed solution by  Algorithm~\ref{alg1} is feasible to {\bf P1}.
Furthermore, we have the following discussions.
\newcounter{qqcounter}
\begin{list}{{\it \arabic{qqcounter})~}}
{\usecounter{qqcounter}
\setlength\leftmargin{0em}
\setlength\labelwidth{0em}
\setlength\labelsep{0em}
\setlength\itemsep{0em}
}
\item In modifying {\bf P1}, we remove the battery capacity constraint \eqref{eqn:Yt bds} and instead impose a new constraint \eqref{eqn:delta_u} on the overall change of battery energy level over $T_o$ slots to be $\Delta_u$. Note that this constraint is set as a desired outcome, \ie $\Delta_u$ is a desired value. The actual solution $\abf_t^*$ in the proposed algorithm may not satisfy this constraint at the end of the $T_o$-slot period, and thus may not be feasible to {\bf P3}. Nonetheless, setting $A_o$ and $V$ as in \eqref{eqn:A_o} and \eqref{eqn:V_max} guarantees that  $\{\abf_t^*\}$ satisfy the battery capacity constraint \eqref{eqn:Yt bds} and therefore are feasible to {\bf P1}.

\item In designing the real-time algorithm by Lyapunov framework, virtual queue $X_t$ in \eqref{eqn:queue X} which we introduce for the average delay constraint \eqref{eqn:avg_x_dt} can only ensure  the constraint is satisfied with a margin as indicated below \eqref{eqn:queue X}.
As a result, constraint \eqref{eqn:avg_x_dt} can only be approximately satisfied. However, this relaxation is mild in practice for the average delay performance. Note that  the per-load maximum delay constraint \eqref{eqn:d_w per time} is strictly satisfied for $d_t^*$ by Algorithm~\ref{alg1}. We will show in simulation that the achieved average delay $\overline{d_w}$ by Algorithm~\ref{alg1} in fact meets constraint \eqref{eqn:avg_x_dt}.
\item We point out that the load scheduling and energy storage control decisions are provided in closed-form by Algorithm~\ref{alg1}. Thus, the algorithm is particularly suitable for real-time implementation with a constant computational complexity $\Oc(1)$. Furthermore, no statistical assumptions on the loads, renewable source, and pricing $\{W_t,S_t,P_t\}$ are required in the algorithm. They can be  non-stochastic or stochastic with arbitrary dynamics (including non-stationary processes). This allows the algorithm to be applied to general scenarios, especially when these statistics are difficult to predict.
Finally, despite that Algorithm~\ref{alg1} is a suboptimal solution for {\bf P1}, we will show in the following that it provides a provable performance guarantee.
\end{list}

\section{Performance Analysis}\label{subsec:PA}
In this section, we analyze the performance of Algorithm~\ref{alg1} and discuss the mismatch involved in some constraints as a result of the real-time algorithm design.
\subsection{Algorithm Performance}
To evaluate the proposed algorithm, we consider a $T$-slot look-ahead problem. Specifically, we partition $T_o$ slots into $T$ frames with $T_o=MT$, for  $M,T\in \mathbb{N}^+$.
For each frame, we consider the same problem as {\bf P1} but the objective is the $T$-slot averaged cost within the frame\ and the constraints are all related to time slots within the frame. In addition, we assume $\{W_t,S_t,P_t\}$ for the entire frame are known beforehand. Thus,  the problem becomes a non-causal static optimization problem and we call it a $T$-slot lookahead problem.   Let   $u_m^\textrm{opt}$ be the corresponding minimum objective value achieved by a $T$-slot look-ahead optimal solution over the $m$th frame.
We intend to bound the performance of Algorithm~\ref{alg1} (with no knowledge of future values of $\{W_t,S_t,P_t\}$) to the optimal $T$-slot lookahead performance (with full knowledge of $\{W_t,S_t,P_t\}$ in a frame).

We denote the objective value of {\bf P1} achieved by Algorithm~\ref{alg1} over $T_o$-slot period by $u^*(V)$, where  $V$ is the weight value used in Algorithm~\ref{alg1}. The following theorem provides a bound of the cost performance under our proposed real-time algorithm to $u_m^\textrm{opt}$ under the  $T$-slot lookahead optimal solution.

\begin{theorem} \label{thm1}
Consider $\{W_t,S_t,P_t\}$ being any arbitrary processes over time. For any $M,T\in \mathbb{N}^+$ satisfying $T_o=MT$,
the $T_o$-slot average system cost under Algorithm~\ref{alg1} is bounded by
\begin{align}\label{thm1:bd}
&u^*(V)-\frac{1}{M}\sum_{m=0}^{M-1}u_m^\textrm{opt} \leq\frac{GT}{V}+ \frac{L(\Thetabf_{0})-L(\Thetabf_{T_o})}{VT_o}\nn\\
&+\frac{C_u'(\Gamma_u)(H_{u,0}-H_{u,T_o})+\alpha C_d'(\Gamma_d)(H_{d,0}-H_{d,T_o})}{T_o}
\end{align}
where $G$ is given in \eqref{eqn:app1_G_value} and the upper bound is finite.
In particular, as $T_o\to \infty$, we have
\begin{align}\label{thm1:bd_longterm}
\lim_{T_o\to \infty}u^*(V)-\lim_{T_o\to \infty}\frac{1}{M}\sum_{m=0}^{M-1}u_m^\textrm{opt}\le \frac{G T}{V}.
\end{align}
\end{theorem}
\IEEEproof
See Appendix \ref{appD}.

\emph{Remark}:  Theorem~\ref{thm1} shows  that our proposed algorithm is able to track the ``ideal" $T$-slot lookahead optimal solution with a bounded gap, for all possible $M$ and $T$.
Also, for the best performance, we should always choose $V=V_{\max}$. The bound in \eqref{thm1:bd_longterm} gives the asymptotic performance as $T_o$ increase. Since $V_{\max}$ in \eqref{eqn:V_max} increases with $B_{\max}$, it follows that Algorithm~\ref{alg1} is asymptotically   equivalent to the optimal $T$-slot look-ahead solution as the battery capacity and $T_o$ go to infinity.
Note that, as $T_o\to\infty$, {\bf P1} becomes an infinite time horizon problem with average sample path cost objective\footnote{As $T_o \to \infty$, constraint \eqref{eqn:delta_u} becomes $\lim_{T_o\rightarrow\infty}\frac{1}{T_o}\sum_{\tau=0}^{T_o-1}(Q_{\tau}+S_{r,\tau}-D_{\tau})=0$ as in the infinite time horizon problem \cite{Li&Dong:acssc2013}.}. The bound in \eqref{thm1:bd_longterm} provides the performance gap of long-term time-averaged sample-path system cost of our proposed algorithm to the $T$-slot look-ahead policy.

\subsection{Design Approximation}
\subsubsection{Average scheduling delay $d^{\max}$}
Recall that, by Algorithm~\ref{alg1}, using the virtual queue $X_t$ in \eqref{eqn:queue X}, the average delay constraint \eqref{eqn:avg_x_dt} is approximately satisfied with a margin, \ie $\overline{d_w}\leq d^{\max}+\epsilon_d$, where $\epsilon_d\triangleq (X_{T_o}-X_0)/T_o$ is the margin. We now bound $\epsilon_d$ below.
\begin{proposition}\label{prop4}
Under Algorithm~\ref{alg1}, the margin $\epsilon_d$ for constraint \eqref{eqn:avg_x_dt} is bounded as follows
\begin{align}\label{eqn:error dmax}
|\epsilon_d|\le \sqrt{\frac{2G}{\mu T_o}+\frac{L(\Thetabf_o)}{\mu T_o}}+\frac{|X_0|}{T_o}.
\end{align}
\end{proposition}
\IEEEproof
See Appendix~\ref{appE}.

Proposition~\ref{prop4} indicates that the margin $\epsilon_d \to 0$ as $T_o\to \infty$. Thus, the average delay is asymptotically satisfied. Note that, for $X_0=0$, $\epsilon_d \ge 0$. If $X_0>0$, it is possible that $\epsilon_d<0$ and constraint \eqref{eqn:avg_x_dt} is satisfied with a negative margin. However, this will drive $d_t^*$ to be smaller which may cause higher system cost. Thus, we set $X_0=0$ in Algorithm~\ref{alg1}.

\subsubsection{Mismatch of $\Delta_u$}
In our design, we set $\Delta_u$ to be a desired value for the change of battery energy level over $T_o$-slot period as in new constraint \eqref{eqn:delta_u}.
This value may not be achieved by Algorithm~\ref{alg1}.
Define the mismatch by $\epsilon_u\triangleq \sum_{\tau=0}^{T_o-1}(Q_\tau+S_{r,\tau}-D_\tau )-\Delta_u$. The bound for $\epsilon_u$  follows the result in \cite[Proposition 3]{Li&Dong:JSAC15} and is shown below.
\begin{align}
|\epsilon_u|\le 2\Gamma_u+R_{\max}+VP_{\max}+VC_u'(\Gamma_u)+D_{\max}.
\end{align}

Note that  $V_{\max}$ in \eqref{eqn:V_max} increases as $|\Delta_u|$ decreases, and a larger $V_{\max}$ is preferred for better performance by Theorem~\ref{thm1}. Thus, a smaller $|\Delta_u|$ is preferred.
Note also that our simulation study shows that the actual mismatch $\epsilon_u$\ is much smaller than this
upper bound.

\section{Simulation Results} \label{sec:sim}
\begin{figure}[t]
\centering
\includegraphics[width=3in]{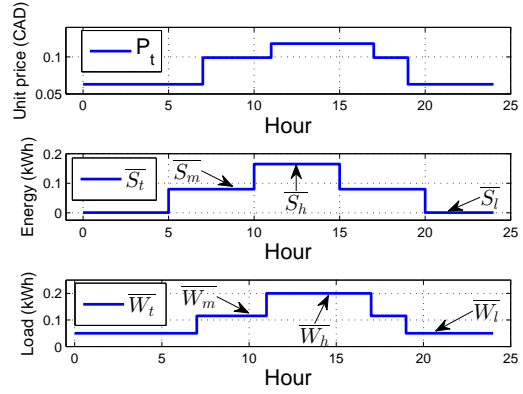}
\caption{\small Mean load $\overline{W}_t$, mean renewable generation $\overline{S}_t$, and real-time price $P_t$ over 24 hours.}
\label{fig:C_S_W_distribution}
\end{figure}
We set each slot to be 5 minutes and consider a 24-hour duration.
Thus, we have $T_o=288$ slots for each day.
We assume $P_t$, $S_t$ and $W_t$ do not change within each slot.
We collect data from Ontario Energy Board \cite{website:Ontario_Elec} to set the price $P_t$. As shown Fig.~\ref{fig:C_S_W_distribution} top, it follows a three-stage price pattern as $\{P_h, P_m, P_l\}=\{\$0.118, \$0.099, \$0.063\}$ and is periodic every 24 hours.
We assume $\{S_t\}$ to be solar photovoltaic energy. It is a non-stationary process, with the mean amount $\overline{S}_t=\mathbb{E}[S_t]$ changing periodically over 24 hours, and having three-stage values as $\{\overline{S}_{h},\overline{S}_{m},\overline{S}_{l}\}=\{1.98,0.96,0.005\}/12$~{kWh} per slot\footnote{We set the mean renewable amount to be $\{\overline{S}_{h},\overline{S}_{m},\overline{S}_{l}\}=\{1.98,0.96,0.005\}$~{kWh} per hour. Converting to per slot energy amount with $5$ minutes duration, we have  $\{1.98,0.96,0.005\}/12$~{kWh} per slot. We set these values based on that the average energy harvested by photovoltaic is about 20 kWh within a day for a residential home.} and standard deviation as $\sigma_{S_i}=0.4\overline{S}_i$, for $i=h,m,l$, as shown in Fig.~\ref{fig:C_S_W_distribution} middle. We assume the load $\{W_t\}$ is  a non-stationary process, having three-stage mean values $\overline{W}_{t}=\mathbb{E}[W_t]$ as $\{\overline{W}_{h},\overline{W}_{m},\overline{W}_{l}\}=\{2.4,1.38,0.6\}/12$~{kWh} per slot\footnote{Similar as $\overline{S}_t$, the mean load is converted from energy amount per hour  to per slot. The mean values are set based on that the each residential household  on average consumes $1\sim 2$~kWh per hour.} with standard deviation as $\sigma_{W_i}=0.2\overline{W}_i$, for $i=h,m,l$, as shown in Fig.~\ref{fig:C_S_W_distribution} bottom. For each load $W_t$, we generate $\lambda_t$ from a uniform distribution  with interval $[1,12]$, and $\rho_t=W_t/\lambda_t$. We set $d_t^{\max}$ in \eqref{eqn:d_w per time} to be identical for all $t$.

We set the battery related parameters as follows: $R_{\max}=D_{\max}=0.165$~{kWh}\footnote{Assuming a household consumes $1 \sim 2$ kWh per hour on average, we set the values of $R_{\max}$ and $D_{\max}$ to be in the same range for load supply. Thus, per slot, we have $R_{\max} = D_{\max} = 1.98 \text{~kWh}/12 = 0.165$ kWh.}, $C_{\textrm{rc}}=C_{\textrm{dc}}=0.001$, $B_{\min}=0$, and the battery initial energy level $B_{0}=0$.
Unless specified, we set $B_{\max}=3$~kWh. We set $E_{\max}=0.3$~{kWh}. Also, we set the weights $\alpha=1$ and $\mu=1$ as the default values.
Since $V_{\max}$ increases as $|\Delta_u|$ decreases, to achieve best performance\footnote{A detailed study of $\Delta_u$ can be found in \cite{Li&Dong:JSAC15}.}, we set $\Delta_u=0$ and $V=V_{\max}$.

We consider an exemplary case where the battery usage cost and the delay cost are both quadratic functions, given by $C_u(\overline{x_u})=k_u\overline{x_u}^2$ and $C_d(\overline{d_w})=k_d\overline{d_w}^2$.
The constant $k_u>0$ is a battery cost coefficient depending on the battery characteristics. We set it as $k_u=0.2$.\footnote{The value of $k_u$ depends on the type of battery and current battery technology. We have tested a range of values for $k_u$ and studied the effect of $k_u$ on the system cost performance in our previous work \cite{Li&Dong:JSAC15}.}
The constant $k_d>0$ is a normalization factor based on the desired maximum average delay $d^{\max}$ in \eqref{eqn:avg_x_dt}. It is set as $k_d={1}/{(d^{\max})^2}$, such that the cost for an average delay $\overline{d_w}=d^{\max}$ is normalized to 1.\footnote{The normalization by $k_d$  ensures the delay cost is properly defined. Intuitively, a longer delay is encouraged in order to allow more energy cost  saving, as long as the desired maximum average delay is satisfied. Thus, the cost  $C_d(\overline{d_w})$ should not be affected by the absolute delay value $\overline{d_w}$, but rather its relative value to $d^{\max}$, \ie ${\overline{d_w}^2}/{(d^{\max})^2}$.}
The optimal $\gamma_{i,t}^*$ in \eqref{eqn:optimal gamma} can be determined with  $C_i'(\Gamma_i)=2k_i\Gamma_i$, and $C_i'^{-1}\left(-\frac{H_{i,t}}{V\beta_i}\right)=-\frac{H_{i,t}}{2\beta_{i} k_iV}$,  for $i=u,d$.

\subsection{An Example of Load Scheduling}
\begin{figure}[t]
\centering
\includegraphics[width=3in]{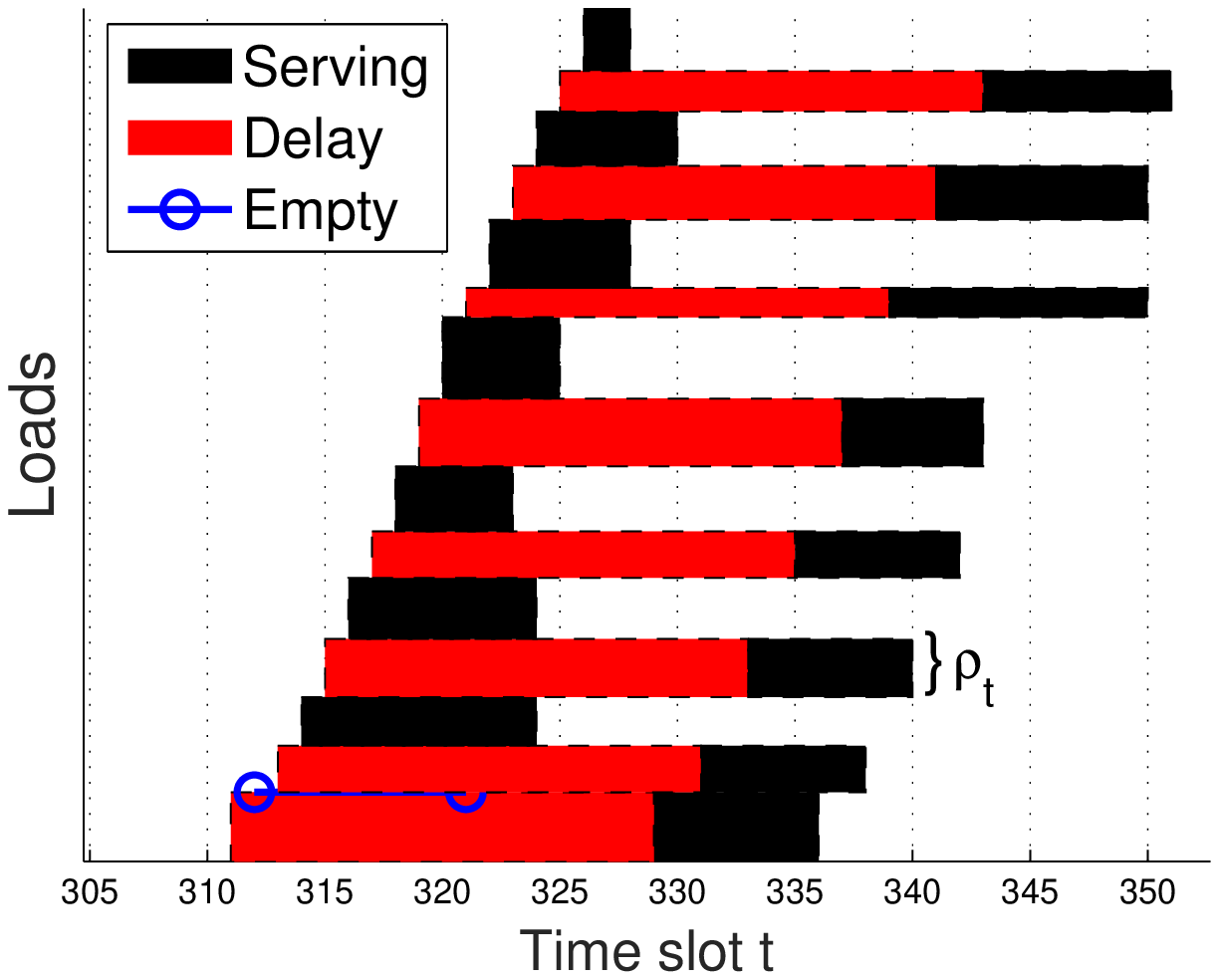}
\caption{\small A trace of load scheduling results over ($d_t^{\max}=d^{\max}=18$, $\alpha=0.005$).}
\label{fig:schedule}
\includegraphics[width=3in]{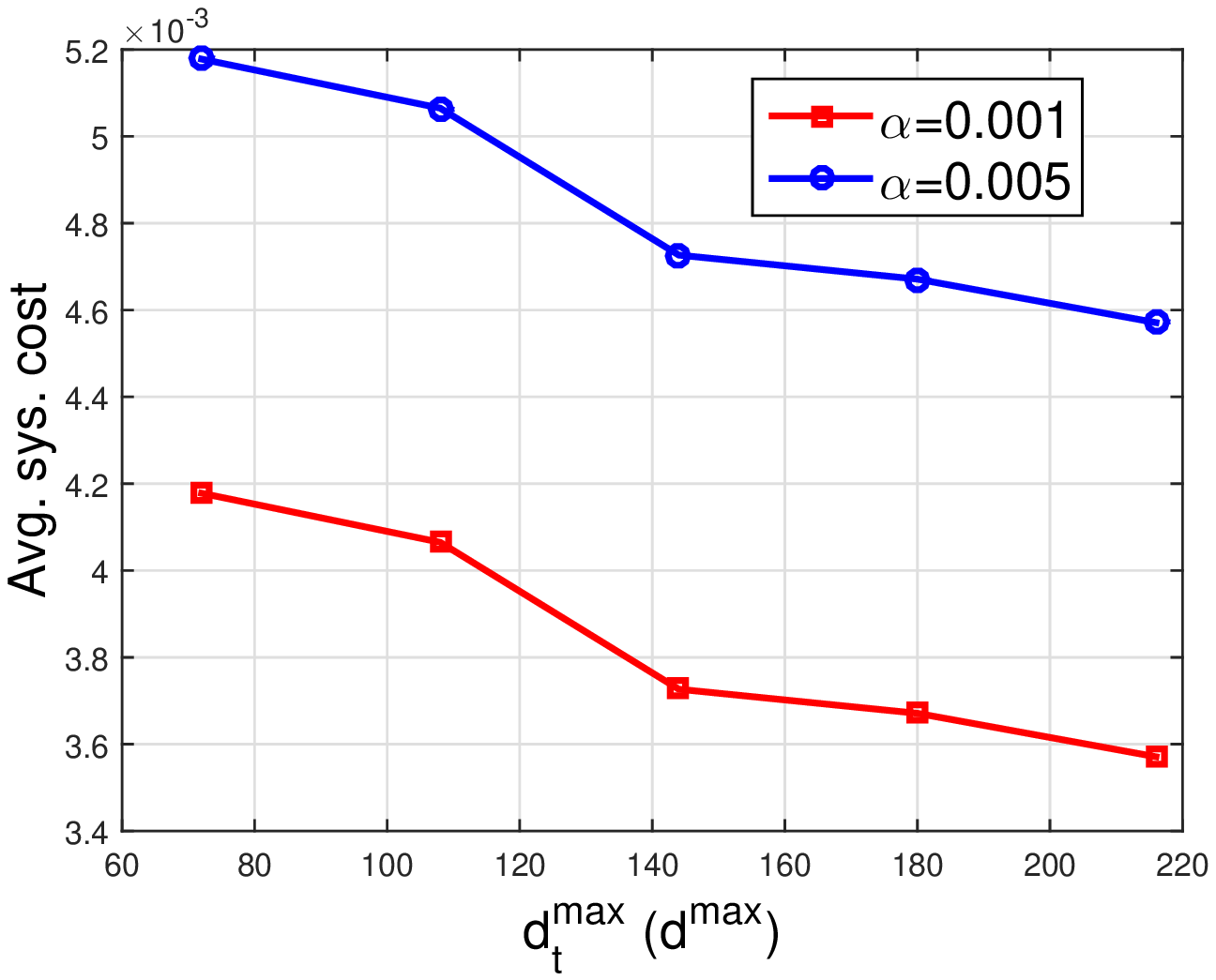}
\caption{\small Average system cost vs. $d_t^{\max}$ ($d^{\max}=d_t^{\max}$).}
\label{fig:SysCost_vs_dmax}
\end{figure}

In Fig.~\ref{fig:schedule}, we show a fraction of the load scheduling results by Algorithm~\ref{alg1}, where we set $d_t^{\max}=d^{\max}=18$ slots and $\alpha=0.005$.\footnote{We set the value of $\alpha$ to ensure that the weighted delay cost  $\alpha C_d(\overline{d_w})$ is comparable to the other two costs in the objective of {\bf P1}. Thus, both delay and energy cost are active factors in making the storage and scheduling decision in simulation.} Each horizontal bar represents a scheduled load $W_t$ with the width representing the load intensity $\rho_{t}$ and the length representing the total duration from arrival to service being completed. For a delayed load, the delay is indicated in different color before the load is scheduled. In this example, we see some loads are immediately scheduled, while others are scheduled at $d_t^{\max}$. The total energy demand at each time slot is the vertical summation over all loads that are scheduled in this time slot.
\begin{figure}[t]
\centering
\includegraphics[width=3in]{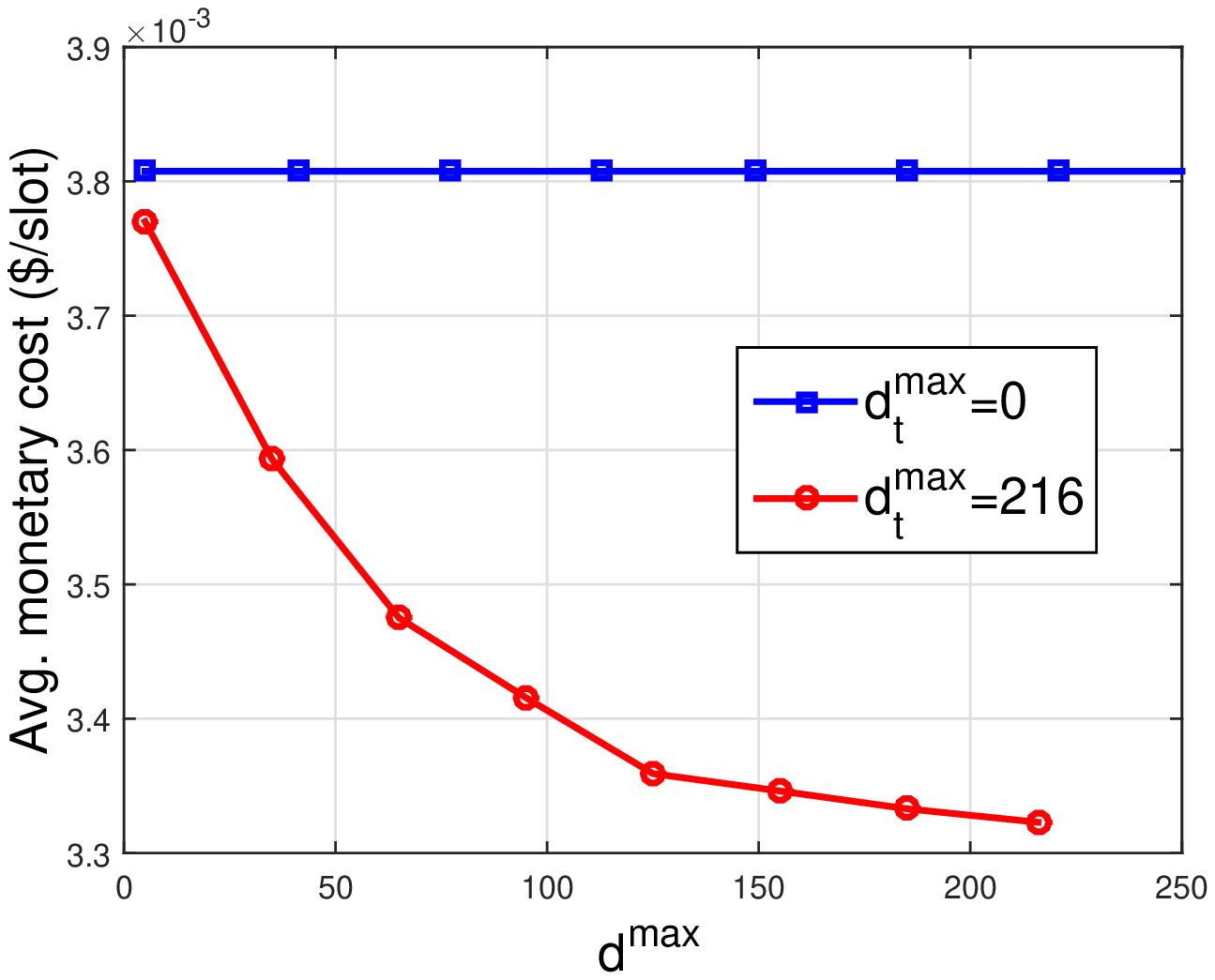}
\caption{\small Monetary cost vs. $d^{\max}$ ($\alpha=0.005$).}
\label{fig:MonetaryCost_vs_dmax}
\includegraphics[width=3in]{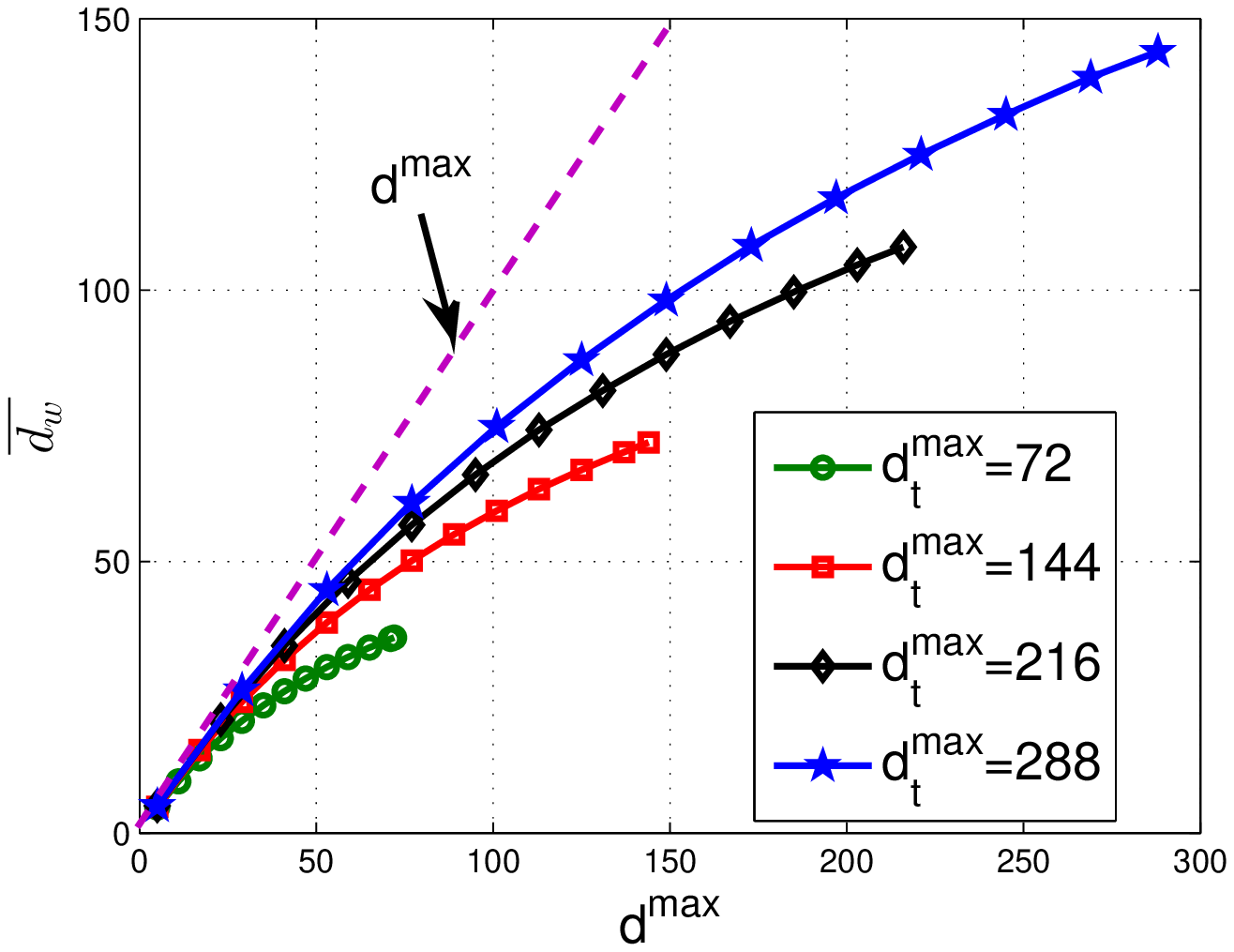}
\caption{\small $\overline{d_w}$ vs. $d^{\max}$ ($\alpha=0.005$).}
\label{fig:AvgDelay_vs_dmax_d_t_max}
\end{figure}
\begin{figure}[t]
\centering
\includegraphics[width=3in]{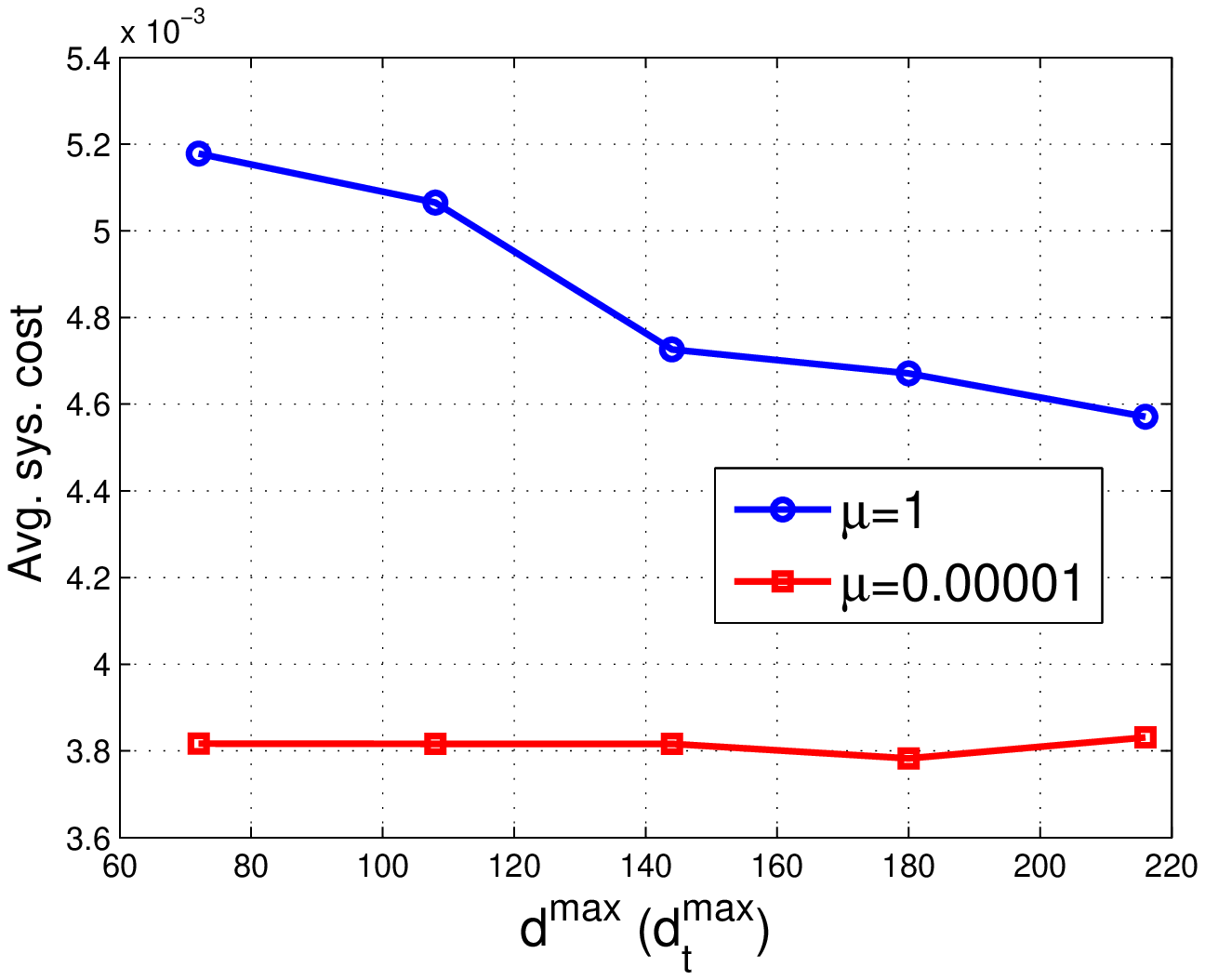}
\caption{\small Average system cost vs. $d^{\max}$ ($d_t^{\max}=d^{\max}$).}
\label{fig:SysCost_vs_dmax_mu_d}
\includegraphics[width=3in]{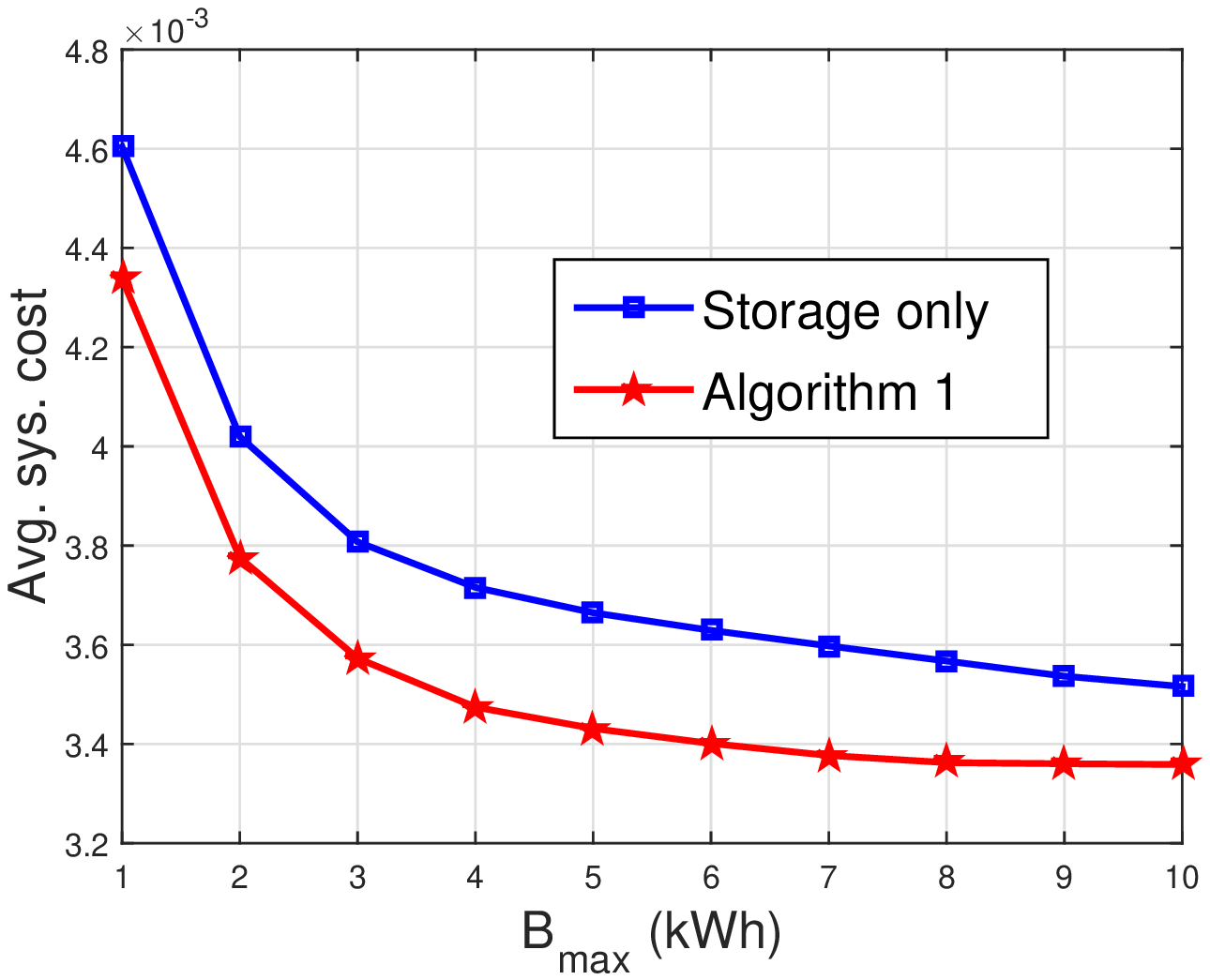}
\caption{\small Average system cost vs. $B_{\max}$ ($\alpha=0.001$).}
\label{fig:SysCost_vs_Bmax}
\end{figure}

\subsection{Effect of Scheduling Delay Constraints}
\subsubsection{Effect of $d_t^{\max}$ and $d^{\max}$}
We study how the average system cost objective of {\bf P1} under our proposed algorithm varies with different delay requirements.  We set $d_t^{\max}=d^{\max}$, $\forall t$, and plot the average system cost vs. $d_t^{\max}$ in Fig.~\ref{fig:SysCost_vs_dmax}, for different values of weight $\alpha$ in the cost objective.
As can be seen, the system cost decreases as $d^{\max}(d_t^{\max})$ increases. This shows that relaxing the average delay constraint gives more flexibility to load scheduling, where each load can be scheduled at lower electricity price, resulting in lower system cost.
This demonstrates that flexible load scheduling is more beneficial to the overall system cost. In addition, we see that a larger value $\alpha$ gives more weight on minimizing the delay in the objective, resulting in a higher system cost.

Next, we study the effect of  load delay constraints on the monetary cost, \ie the cost of energy purchasing and battery degradation, given by $\overline{J}+\overline{x_e}+ C_u(\overline{x_u})$ in the objective of {\bf P1}.
The monetary cost indicates how much saving a consumer could actually have, by allowing longer service delay.  In Fig.~\ref{fig:MonetaryCost_vs_dmax}, we plot the monetary cost vs. $d^{\max}$ for  $d_t^{\max}=216$.
We see a clear trade-off between the monetary cost and the load delay. The trade-off curve can be used to determine the desired operating point. For comparison, we also consider the case in which all loads are served immediately after arrival, \ie  $d_t^{\max}=0$. This is essentially the case with only energy storage but no load scheduling. Thus, the monetary cost is independent of $d^{\max}$. We see a substantial gap between  the two curves and the gap increases with $d^{\max}$. This clearly shows the benefit of joint load scheduling and energy storage management.

\subsubsection{Average delay $\overline{d_w}$}
We now study the average delay $\overline{d_w}$ achieved by our proposed algorithm vs. $d^{\max}$ for various $d^{\max}_t$ in Fig.~\ref{fig:AvgDelay_vs_dmax_d_t_max}.
As we see, the actual averaged delay $\overline{d_w}$ increases with the average delay $d^{\max}$ requirement. This is because, with a more relaxed constraint on the  delay, loads can be shifted to a later time in order to reduce the system cost, resulting in larger average delay. However, the increase is sublinear with respect to $d^{\max}$.
Similarly, we observe that increasing the per load maximum  delay $d_t^{\max}$  increases $\overline{d_w}$.  Finally, recall that we study the margin  for average delay constraint \eqref{eqn:avg_x_dt} under our proposed algorithm in Proposition~\ref{prop4}. To see how the resulting $\overline{d_w}$ meets constraint \eqref{eqn:avg_x_dt}, we plot the line $d^{\max}$ in Fig.~\ref{fig:AvgDelay_vs_dmax_d_t_max}. As we see, $\overline{d_w}$ is below $d^{\max}$ for all values of $d^{\max}$ and $d_t^{\max}$.
\subsubsection{Effect of $\mu$} Weight $\mu$ is used to control the relative importance of virtual queues related to the battery and those to delay in Lyapunov function $L(\Thetabf_t)$ in \eqref{eqn:lyapunov function} and  Lyapunov drift $\Delta(\Thetabf_t)$. For Lyapunov drift $\Delta(\Thetabf_t)$, if $\mu$ is large, the two queues $X_t$ and $H_{d,t}$ related to the load delay will dominate the drift. This will affect the drift-plus-cost objective considered in our proposed algorithm and thus the performance. To study the effect of $\mu$ on the performance, in Fig.~\ref{fig:SysCost_vs_dmax_mu_d}, we evaluate the average system cost for different values of $\mu$.
 We see that
a lower system cost is achieved by smaller value of $\mu$.
This is because, with smaller $\mu$, the drifts of $X_t$ related to delay is less significant in the overall drift. This allows wider difference between  $\overline{d_w}$ and $d^{\max}$, \ie smaller  $\overline{d_w}$ and lower delay cost.

\subsection{Performance  vs. Battery Capacity}
We consider two other algorithms for comparison:
A) \emph{No storage or scheduling}: In this case, neither energy storage nor load scheduling is considered. Each load is served immediately using energy purchased from the conventional grid and/or renewable generator.
B) \emph{Storage only}: In this method, only battery storage is considered but every load is served immediately without a delay. This is essentially the algorithm provided in \cite{Li&Dong:JSAC15}.

In Fig.~\ref{fig:SysCost_vs_Bmax}, we compare our proposed algorithm to the above two alternative algorithms under various battery capacity $B_{\max}$.
Since  algorithm~A does not use a battery, the system cost is unchanged over $B_{\max}$ and is 0.01. We do not plot the curve as the cost is much higher than the rest of two algorithms we considered.
For algorithm~B and our proposed Algorithm~\ref{alg1}, as can be seen, the system costs reduces as $B_{\max}$ increases.
This is because a larger battery capacity allows charging/discharging to be more flexible based on the current demand and electricity price, resulting in a lower system cost.
Comparing the two, we see that joint load scheduling and energy storage control provides further reduction in system cost.

\section{Conclusion}\label{sec:conclusion}
In this work, we have considered joint energy storage management and load scheduling for the ESM system, where renewable source, loads, and price may be non-stationary and their statistics are unknown. For load scheduling, we have characterized each load task  by its power intensity and service duration, and have considered the maximum per load  delay and maximum average delay requirements. For storage control, our storage model includes details of the battery operational constraints and cost. Aiming at minimizing the overall system cost over a finite period of time, we have designed a real-time algorithm for joint load scheduling and energy storage control, where we have provided a closed-form per slot scheduling and energy storage decisions. As a result, we have shown that the joint load scheduling and energy storage control can in fact be separately and sequentially determined in our real-time algorithm.
Furthermore, we have shown that our proposed real-time algorithm has a bounded performance guarantee from an optimal $T$-slot look-ahead solution and is asymptotically equivalent to the optimal $T$-slot look-ahead solution, as the battery capacity and time period go to infinity.
Simulation results have demonstrated the
gain of joint load scheduling and storage control provided by our proposed algorithm  over other real-time schemes which consider neither storage nor scheduling, or with storage only.

\appendices

\section{Proof of Equivalence of {\bf P2} and {\bf P3}}\label{appA}
\IEEEproof
The proof follows the same approach as in  \cite[Lemma 1]{Li&Dong:JSAC15}.
Let $u^o_2$ and $u^o_3$ denote the minimum objective values of {\bf P2} and {\bf P3}, respectively.
Since the optimal solution of \textrm{\bf P2} satisfies all constraints of \textrm{\bf P3}, it is a feasible solution of {\bf P3}.
Thus, we have $u^o_3\le u^o_2$.
By Jensen's inequality and convexity of $C_i(\cdot)$ for $i=d,u$, we have
$\overline{C_d(\gamma_d)}\geq C_d(\overline{\gamma_d})=C_d(\overline{d_w})$
and $\overline{C_u(\gamma_u)}\geq C_u(\overline{\gamma_u})=C_u(\overline{x_u})$. This means $u^o_3\ge u^o_2$.
Hence, we have $u^o_2=u^o_3$ and \textrm{\bf P3} and \textrm{\bf P2} are equivalent.
\endIEEEproof

\section{Upper Bound on Drift-Plus-Cost Function}\label{appB}
The following lemma presents an upper bound on the drift $\Delta(\Thetabf_t)$.
\begin{lemma}\label{lemma2}
The one-slot Lyapunov drift $\Delta(\Thetabf_t)$ is upper bounded by
\allowdisplaybreaks
\begin{align}\label{eqn:lemma drift bound}
&\Delta(\Thetabf_t) \leq  Z_t\left(E_t+S_{r,t}+S_{w,t}-\sum_{\tau=0}^{t}\rho_{\tau}1_{S,t}(d_\tau)-\frac{\Delta_u}{T_o}\right) \nn\\
&+H_{u,t}\gamma_{u,t}-H_{u,t}(E_t+S_{r,t})+\mu X_t(d_t-d^{\max})+G\nn\\
&-|H_{u,t}|\left(S_{w,t}-\sum_{\tau=0}^{t}\rho_{\tau}1_{S,t}(d_\tau)\right)+\mu H_{d,t}(\gamma_{d,t}-d_t)
\end{align}
where
$G=\frac{1}{2}\max\left\{\left(R_{\max}-\frac{\Delta_u}{T_o}\right)^2,\left(D_{\max}+\frac{\Delta_u}{T_o}\right)^2\right\} +\frac{1}{2}\max\{R^2_{\max},D^2_{\max}\}
+\frac{\mu}{2}\max\left\{(d^{\max})^2,(d_t^{\max}-d^{\max})^2\right\}
\allowbreak+\frac{\mu}{2}(d_t^{\max})^2$.
\end{lemma}
\IEEEproof
From the definition of $\Delta(\Thetabf_t)$, we have
\begin{align}\label{eqn:drift part1}
\Delta(\Thetabf_t)
&\triangleq L(\Thetabf_{t+1})-L(\Thetabf_{t}) \nn\\
&=\frac{1}{2}[Z^2_{t+1}-Z^2_t+(H_{u,t+1}^2-H_{u,t}^2)]\nn\\
&\quad+\frac{\mu}{2}\left[(X^2_{t+1}-X^2_t+H_{d,t+1}^2-H_{d,t}^2)\right]
\end{align}
where from queue \eqref{eqn:queue Z}, $Z^2_{t+1}-Z^2_t$ can be presented by
\begin{align}\label{eqn:drift z}
\frac{Z_{t+1}^2-Z_t^2}{2}&=Z_t\left(Q_t+S_{r,t}-D_t-\frac{\Delta_u}{T_o}\right)\nn\\
&\quad\quad+\frac{(Q_t+S_{r,t}-D_t-\frac{\Delta_u}{T_o})^2}{2}.
\end{align}
Note that from \eqref{eqn:delta_u}, we have $\frac{\Delta_u}{T_o}\le \max\{R_{\max},D_{\max}\}$.
For a given value of $\Delta_u$, by \eqref{eqn:rc_bds} and \eqref{eqn:dc_bds}, $(Q_t+S_{r,t}-D_t-\frac{\Delta_u}{T_o})^2$ is upper bound by $\max\left\{(R_{\max}-\frac{\Delta_u}{T_o})^2,(D_{\max}+\frac{\Delta_u}{T_o})^2\right\}$.
By the supply-demand balance \eqref{eqn:Wt_constraint}, the first term on RHS of \eqref{eqn:drift z} can be replaced by
\allowdisplaybreaks
\begin{align}\label{eqn:z replace}
&Z_t\left(Q_t+S_{r,t}-D_t-\frac{\Delta_u}{T_o}\right)\nn\\
&=Z_t\left(E_t+S_{r,t}+S_{w,t}-\sum_{\tau=0}^{t}\rho_{\tau}1_{S,t}(d_\tau)-\frac{\Delta_u}{T_o}\right).
\end{align}

From \eqref{eqn:queue H1}, $H_{u,t+1}^2-H_{u,t}^2$ in \eqref{eqn:drift part1} can be presented by
\begin{align}\label{eqn:Hu part1}
&\frac{H_{u,t+1}^2-H_{u,t}^2}{2}=H_{u,t}(\gamma_{u,t}-x_{u,t})+\frac{(\gamma_{u,t}-x_{u,t})^2}{2}.
\end{align}
Note that from \eqref{eqn:x_u} and \eqref{eqn:gamma_bds}, the second term of RHS in \eqref{eqn:Hu part1} is upper bounded by $(\gamma_{u,t}-x_{u,t})^2\leq \max\{R_{\max}^2,D_{\max}^2\}$.

We now find the upper bound for $-H_tx_{u,t}$ in the first term on RHS of \eqref{eqn:Hu part1}.
By the supply-demand balance \eqref{eqn:Wt_constraint}, $-H_tx_{u,t}$ can be replaced by
\begin{align}\label{eqn:Hu part2}
-H_tx_{u,t}&=-H_{u,t}(|Q_t+S_{r,t}-D_t|)\nn\\
&\hspace*{-2.3em}=-H_{u,t}\left(\left|E_t+S_{r,t}+S_{w,t}-\sum_{\tau=0}^{t}\rho_{\tau}1_{S,t}(d_\tau)\right|\right).
\end{align}
The upper bound of $-H_tx_{u,t}$ in \eqref{eqn:Hu part2} is obtained as follows.

\newcounter{q1counter}
\begin{list}{{\it \arabic{q1counter}$\left.\right)$~}}
{\usecounter{q1counter}
\setlength\leftmargin{0em}
\setlength\labelwidth{0em}
\setlength\labelsep{0em}
\setlength\itemsep{0em}
}
\item {\it For $H_{u,t}\geq0$}: We have
\begin{align}
&-H_{u,t}\left(\left|E_t+S_{r,t}+S_{w,t}-\sum_{\tau=0}^{t}\rho_{\tau}1_{S,t}(d_\tau)\right|\right)\nn\\
&\leq-H_{u,t}\left(E_t+S_{r,t}+S_{w,t}-\sum_{\tau=0}^{t}\rho_{\tau}1_{S,t}(d_\tau)\right)\nn\\
&=-H_{u,t}(E_t+S_{r,t})-H_{u,t}\left(S_{w,t}-\sum_{\tau=0}^{t}\rho_{\tau}1_{S,t}(d_\tau)\right).\nn
\end{align}
\item {\it For $H_{u,t}<0$}: We have
\begin{align}
&-H_{u,t}\left(\left|E_t+S_{r,t}+S_{w,t}-\sum_{\tau=0}^{t}\rho_{\tau}1_{S,t}(d_\tau)\right|\right)\nn\\
&\leq-H_{u,t}\left(\left|E_t+S_{r,t}\right|+\left|S_{w,t}-\sum_{\tau=0}^{t}\rho_{\tau}1_{S,t}(d_\tau)\right|\right)\nn\\
&\leq-H_{u,t}\left(E_t+S_{r,t}+\sum_{\tau=0}^{t}\rho_{\tau}1_{S,t}(d_\tau)-S_{w,t}\right)\nn\\
&=-H_{u,t}(E_t+S_{r,t})+H_{u,t}\left(S_{w,t}-\sum_{\tau=0}^{t}\rho_{\tau}1_{S,t}(d_\tau)\right).\nn
\end{align}
Combine the above cases for $H_{u,t}$, we have $-H_tx_{u,t}$ in \eqref{eqn:Hu part2} upper bounded by
\allowdisplaybreaks\begin{align}
&-H_{u,t}\left(\left|E_t+S_{r,t}+S_{w,t}-\sum_{\tau=0}^{t}\rho_{\tau}1_{S,t}(d_\tau)\right|\right)\nn\\
&\leq
-H_{u,t}\left(E_t+S_{r,t}\right)-|H_{u,t}|\left(S_{w,t}-\sum_{\tau=0}^{t}\rho_{\tau}1_{S,t}(d_\tau)\right).\nn
\end{align}
\end{list}

From \eqref{eqn:queue X}, we have $X_{t+1}^2\leq \left(X_t+d_{t}-d^{\max}\right)^2$. Thus, $(X^2_{t+1}-X^2_t)$ in \eqref{eqn:drift part1} is bounded by
\begin{align}\label{eqn: X upper bound}
\frac{X_{t+1}^2-X_t^2}{2}&\leq X_t\left(d_{t}-d^{\max}\right)+\frac{1}{2}\left(d_{t}-d^{\max}\right)^2
\end{align}
where by \eqref{eqn:d_w per time}, the last term in RHS of \eqref{eqn: X upper bound} is upper bounded by $(d_{t}-d^{\max})^2\leq \max\left\{\left(d^{\max}\right)^2,\left(d_t^{\max}-d^{\max}\right)^2\right\}$.

From \eqref{eqn:queue H2}, $H_{d,t+1}^2-H_{d,t}^2$ in \eqref{eqn:drift part1} can be presented by
\begin{align}\label{eqn:H2 upper bound}
\frac{H_{d,t+1}^2-H_{d,t}^2}{2}&=H_{d,t}(\gamma_{d,t}-d_{t})+\frac{(\gamma_{d,t}-d_{t})^2}{2}\nn\\
&\leq H_{d,t}(\gamma_{d,t}-d_{t})+\frac{1}{2}(d_t^{\max})^2
\end{align}
where the last inequality is derived from the bounds of $\gamma_d$ in \eqref{avg_gamma2b} and $d_t$ in \eqref{eqn:d_w per time}.
We give the upper bond of \eqref{eqn:drift part1} as follows
\begin{align}\label{eqn:app1_delta1}
&\Delta(\Thetabf_t)
\triangleq L(\Thetabf_{t+1})-L(\Thetabf_{t})\nn\\
&\leq Z_t\left(E_t+S_{r,t}+S_{w,t}-\sum_{\tau=0}^{t}\rho_{\tau}1_{S,t}(d_\tau)-\frac{\Delta_u}{T_o}\right)\nn\\
&-|H_{u,t}|\left(S_{w,t}-\sum_{\tau=0}^{t}\rho_{\tau}1_{S,t}(d_\tau)\right)+\mu H_{d,t}(\gamma_{d,t}-d_{t})+G\nn\\
&+H_{u,t}\gamma_{u,t}-H_{u,t}(E_t+S_{r,t})+\mu X_t\left(d_{t}-d^{\max}\right)
\end{align}
where $G$ includes all constant terms from the upper bounds of \eqref{eqn:drift z}, \eqref{eqn:Hu part1}, \eqref{eqn: X upper bound} and \eqref{eqn:H2 upper bound}, and is defined as
\begin{align}\label{eqn:app1_G_value}
G&\triangleq\frac{1}{2}\max\left\{\left(R_{\max}-\frac{\Delta_u}{T_o}\right)^2,\left(D_{\max}+\frac{\Delta_u}{T_o}\right)^2\right\}\nn\\ &\quad+\frac{1}{2}\max\{R^2_{\max},D^2_{\max}\}+\frac{\mu}{2}(d_{t}^{\max})^2\nn\\
&\quad+\frac{\mu}{2}\max\left\{(d^{\max})^2,(d_t^{\max}-d^{\max})^2\right\}.
\end{align}
\endIEEEproof

\section{Proof of Proposition~\ref{prop dt}}\label{app:prop dt}
\IEEEproof
To determine the optimal scheduling delay $d_t^*$, we need to compare the objective values of ${\bf \text{\bf P4}_{a1}}$ under all serving options. The optimal delay $d_t^*$ is the one that achieves the minimum objective value. Because $d_t\cdot1_{S,t}(d_t)=0$, we have the following cases:
\begin{enumerate}
\item If the load is immediately served, we have $1_{S,t}(d_t)=1$ and $d_t=0$. The objective value becomes $\omega_o$;
\item If the load is delayed, we have $d_t>0$ and $1_{S,t}(d_t)=0$. The objective function is reduced to $\mu d_t(X_t-H_{d,t})$.
For $X_t-H_{d,t}\geq0$, the objective value is $\omega_1$; Otherwise, the value is $\omega_{d_t^{\max}}$.
\end{enumerate}
Comparing $\omega_o$ to $\omega_1$ or $\omega_{d_t^{\max}}$, we obtain $d_t^*$.
\endIEEEproof

\section{Proof of Lemma~\ref{lemma33}}\label{app:prop1}
\IEEEproof
Since $\mu$, $\alpha$ and $V$ are all positive weights, and $C_i(\gamma_t)$'s are assumed to be continuous, convex and non-decreasing functions with respect to $\gamma_{i,t}$ with maximum derivatives $C_i'(\Gamma_i)<\infty$, for $i=d,u$, the optimal $\gamma^*_{i,t}$'s  are determined by examining the derivatives of the objective functions of ${\bf \text{\bf P4}_{a2}}$ and ${\bf \text{\bf P4}_{b1}}$.
Note that, given $\gamma_{u,t}$\ in \eqref{eqn:gamma_bds} and $\gamma_{d,t}$ in \eqref{avg_gamma2b}, $C_i'(\gamma_{i,t})\ge 0$ and increases with $\gamma_{i,t}$ for $i=d,u$.
For $\beta_i$ defined in Lemma~\ref{lemma33}, we have
\newcounter{q2counter}
\begin{list}{{\it \arabic{q2counter}$\left.\right)$~}}
{\usecounter{q2counter}
\setlength\leftmargin{0em}
\setlength\labelwidth{0em}
\setlength\labelsep{0em}
\setlength\itemsep{0em}
}
\item {\it For $H_{i,t}\geq0$}: We have $\mu_dH_{d,t}+V\alpha C_d'(\gamma_{d,t})>0$ and $H_{u,t}+VC_u'(\gamma_{u,t})>0$. Thus, the objectives of ${\bf \text{\bf P4}_{a2}}$ and ${\bf \text{\bf P4}_{b1}}$ are both monotonically increasing functions, and the minimum values are obtained with $\gamma^*_{i,t}=0$ for $i=d,u$.
\item {\it For $H_{i,t}<-V\beta C_i'(\Gamma_i)$}: Since $VC_i'(\Gamma_i)\ge VC_i'(\gamma_{i,t})$ for $i=d,u$, we have $\mu_dH_{d,t}+V\alpha C_d'(\gamma_{d,t})<0$ and $H_{u,t}+VC_u'(\gamma_{u,t})<0$.
    The objectives of ${\bf \text{\bf P4}_{a2}}$ and ${\bf \text{\bf P4}_{b1}}$ are both monotonically decreasing functions.
    From \eqref{avg_gamma2b}, the minimum objective value of ${\bf \text{\bf P4}_{a2}}$ is reached with $\gamma^*_{d,t}=\Gamma_d$ where $\Gamma_d\triangleq\min\{d_t^{\max},d^{\max}\}$;
    The minimum objective value of ${\bf \text{\bf P4}_{b1}}$ is reached with $\gamma^*_{u,t}=\Gamma_u$ where by \eqref{eqn:gamma_bds}, we have $\Gamma_u\triangleq\min\{R_{\max},D_{\max}\}$;
\item  {\it For $-V\beta_{i} C_i'(\Gamma_i)\le H_{i,t}\le 0$}: In this case, $\gamma^*_{d,t}$ and $\gamma^*_{u,t}$ are the roots of $\mu_dH_{d,t}+V\alpha C_d'(\gamma_{d,t})=0$ and $H_{u,t}+VC_u'(\gamma_{u,t})=0$, respectively. We have $\gamma^*_{i,t}=C_i'^{-1}\left(-\frac{H_{i,t}}{V\beta_{i}}\right)$ for $i=d,u$.
\end{list}

Thus, we have $\gamma^*_{i,t}$ for $i=d,u$ as in \eqref{eqn:optimal gamma}.
\endIEEEproof

\section{Proof of Theorem~\ref{thm1}}\label{appD}
\IEEEproof
A $T$-slot sample path Lyapunov drift is defined by $\Delta_T(\Thetabf_t)\triangleq L(\Thetabf_{t+T})-L(\Thetabf_{t})$. We upper bound it as follows
\allowdisplaybreaks
\begin{align}\label{eqn:app3_Delta_T}
&\Delta_T(\Thetabf_t)
=\frac{Z^2_{t+T}-Z^2_{t}+\left(H^2_{u,t+T}-H^2_{u,t}\right)}{2}\nn\\
& \hspace*{5em} +\frac{\mu\left(X^2_{t+T}-X^2_{t}+H^2_{d,t+T}-H^2_{d,t}\right)}{2}\nn\\
&\leq Z_t\sum_{\tau=t}^{t+T-1}\left( Q_\tau+S_{r,\tau}-D_\tau-\frac{\Delta_u}{T_o}\right)\nn\\
&+\frac{1}{2}\left[\sum_{\tau=t}^{t+T-1}\left(Q_\tau+S_{r,\tau}-D_\tau-\frac{\Delta_u}{T_o}\right)\right]^2\nn\\
&+H_{u,t}\sum_{\tau=t}^{t+T-1}(\gamma_{u,\tau}-x_{u,\tau})+\frac{1}{2}\left[\sum_{\tau=t}^{t+T-1}(\gamma_{u,\tau}-x_{u,\tau})\right]^2\nn\\
&+X_t\sum_{\tau=t}^{t+T-1}(d_{\tau}-d^{\max})+\frac{\mu}{2}\left[\sum_{\tau=t}^{t+T-1}(d_{t}-d^{\max})\right]^2\nn\\
&+H_{d,t}\sum_{\tau=t}^{t+T-1}(\gamma_{d,\tau}-x_{d,\tau})
+\frac{\mu}{2}\left[\sum_{\tau=t}^{t+T-1}(\gamma_{d,\tau}-x_{d,\tau})\right]^2\nn\\
&\leq Z_t\sum_{\tau=t}^{t+T-1}\left(Q_\tau+S_{r,\tau}-D_\tau-\frac{\Delta_u}{T_o}\right)\nn\\
&+H_{u,t}\sum_{\tau=t}^{t+T-1}(\gamma_\tau-x_{u,\tau})+X_t\sum_{\tau=t}^{t+T-1}(d_{\tau}-d^{\max})\nn\\
&+H_{d,t}\sum_{\tau=t}^{t+T-1}(\gamma_{d,\tau}-x_{d,\tau})+G T^2
\end{align}
where $G$ is defined in Lemma \ref{lemma2}.

Assume $T_o=MT$. We consider a per-frame optimization problem below,
with the objective of minimizing the time-averaged system cost within the $m$th frame of length $T$ time slots.
\begin{align}
{\bf P_f:} &\min_{\{\abf_t,\gamma_t\}}
\frac{1}{T}\sum_{t=mT}^{(m+1)T-1}\hspace*{-1em}\left[E_tP_t+x_{e,t}+C_u(\gamma_{u,t})+\alpha C_d(\gamma_{d,t})\right]\nn\\
\rm{s.t} \;\;
&\eqref{eqn:d_w per time},\eqref{equ:Pt_constraint},\eqref{eqn:S2-bounds},\eqref{eqn:rc_dc_constr},
\eqref{eqn:Wt_constraint}-\eqref{eqn:dc_bds_strict},\eqref{eqn:gamma_bds}-\eqref{eqn:gamma2_bds}.\nn
\end{align}
We show that ${\bf P_f}$ is equivalent to ${\bf P1}$ in which $T_o$ is replaced by $T$.
Let $u_m^f$ denote the minimum objective value of ${\bf P_f}$. The optimal solution of ${\bf P1}$ satisfies all constraints of ${\bf P_f}$ and therefore is feasible to ${\bf P_f}$.
Thus, we have $u_m^f\leq u_m^\textrm{opt}$.
By Jensen's inequality and convexity of $C_i(\cdot)$ for $i=d,u$, we have $\overline{C_d(\gamma_d)}\geq C_d(\overline{\gamma_d})=C_d(\overline{d_w})$ and $\overline{C_u(\gamma_u)}\geq C_u(\overline{\gamma_u})=C_u(\overline{x_u})$.
Note that introducing the auxiliary variables $\gamma_{u,t}$ with constraints \eqref{eqn:gamma_bds} and \eqref{avg_r=avg_x}, and $\gamma_{d,t}$ with constraints \eqref{avg_gamma2b} and \eqref{eqn:gamma2_bds} does not modify the problem.
This means $u_m^f\ge u_m^\textrm{opt}$.
Hence, we have $u_m^f= u_m^\textrm{opt}$ and ${\bf P_f}$ and ${\bf P1}$ are equivalent.

From \eqref{eqn:app3_Delta_T} and the objective of ${\bf P_f}$, we have the $T$-slot drift-plus-cost metric for the $m$th frame upper bounded by
\begin{align}\label{eqn:T slot drift plus penalty}
&\hspace*{-.5em}\Delta_T(\Thetabf_t)+V\hspace*{-1em}\sum_{t=mT}^{(m+1)T-1}\hspace*{-1em}[E_t P_t+x_{e,t}+C_u(\gamma_{u,t})+\alpha C_d(\gamma_{d,t})]\nn\\
&\hspace*{-1em}\leq Z_{t}\hspace*{-1em}\sum_{t=mT}^{(m+1)T-1}\hspace*{-.5em}\left(Q_t+S_{r,t}-D_t-\frac{\Delta_u}{T_o}\right)+X_t\hspace*{-.5em}\sum_{\tau=t}^{t+T-1}\hspace*{-.5em}(d_{\tau}-d^{\max})\nn\\
&+H_{u,t}\hspace*{-.5em}\sum_{\tau=t}^{t+T-1}\hspace*{-.5em}(\gamma_{u,\tau}-x_{u,\tau})+G T^2+H_{d,t}\hspace*{-.5em}\sum_{\tau=t}^{t+T-1}\hspace*{-.5em}(\gamma_{d,\tau}-x_{d,\tau})\nn\\
& +V\hspace*{-1em}\sum_{t=mT}^{(m+1)T-1}\hspace*{-1em}\left[E_t P_t+x_{e,t}+C_u(\gamma_{u,t})+\alpha C_d(\gamma_{d,t})\right].
\end{align}
Let $\{\tilde{\pibf}_t\}$ denote a set of feasible solutions of ${\bf P_f}$, satisfying the following relations
\begin{align}
&\sum_{t=mT}^{(m+1)T-1}\left(\tilde{Q}_t+\tilde{S}_{r,t}\right)=\sum_{t=mT}^{(m+1)T-1}\left(\tilde{D}_t+\frac{\Delta_u}{T_o}\right)\label{eqn:avg_rc_dc_bds_per_frame_1}\\
&\sum_{t=mT}^{(m+1)T-1}\tilde{\gamma}_{i,t}=\sum_{t=mT}^{(m+1)T-1}\tilde{x}_{i,t},\ \textrm{for}\  i=u,d\label{eqn:avg_rc_dc_bds_per_frame_2}\\
&\sum_{t=mT}^{(m+1)T-1}\tilde{d}_{t} \le \sum_{t=mT}^{(m+1)T-1}d^{\max}\label{eqn:avg_rc_dc_bds_per_frame_3}
\end{align}
with the corresponding objective value denoted as $\tilde{u}^f_m$.

Note that comparing with ${\bf P1}$, we impose per-frame constraints \eqref{eqn:avg_rc_dc_bds_per_frame_1}-\eqref{eqn:avg_rc_dc_bds_per_frame_3} as oppose to \eqref{eqn:delta_u}, \eqref{avg_r=avg_x}, \eqref{eqn:gamma2_bds} and \eqref{eqn:avg_x_dt} for the $T_o$-slot period, respectively.
Let $\delta\geq 0$ denote the gap of $\tilde{u}^f_m$ to the optimal objective value $u_m^\textrm{opt}$, \ie $\tilde{u}^f_m=u_m^\textrm{opt}+\delta$.

Among all feasible control solutions satisfying \eqref{eqn:avg_rc_dc_bds_per_frame_1}-\eqref{eqn:avg_rc_dc_bds_per_frame_3}, there exists a solution which leads to $\delta\rightarrow0$. The upper bound in \eqref{eqn:T slot drift plus penalty} can be rewritten as
\begin{align}\label{eqn:T slot drift plus penalty, per frame}
&\Delta_T(\Thetabf_t)+V\left[\sum_{t=mT}^{(m+1)T-1}\hspace*{-1em}\left[E_t P_t+x_{e,t}+C_u(\gamma_{u,t})+\alpha C_d(\gamma_{d,t})\right]\right]\nn\\
&\leq G T^2+VT\lim_{\delta\rightarrow0}\left(u_m^\textrm{opt}+\delta\right)= G T^2+VTu_m^\textrm{opt}.
\end{align}
Summing both sides of \eqref{eqn:T slot drift plus penalty, per frame} over $m$ for $m=0,\ldots, M-1$, and dividing them by $VMT$, we have
\begin{align}\label{eqn:T slot drift plus penalty, 2}
&\frac{1}{MT}\sum_{m=0}^{M-1}\sum_{t=mT}^{(m+1)T-1}\left[E_t P_t+x_{e,t}+C_u(\gamma_{u,t})+\alpha C_d(\gamma_{d,t})\right]\nn\\
&\quad+\frac{L(\Thetabf_{T_o})-L(\Thetabf_0)}{VMT}
\leq
\frac{G T}{V}+\frac{1}{M}\sum_{m=0}^{M-1}u_m^\textrm{opt}.
\end{align}

Since $C_i(\overline{\gamma_i})\leq\overline{C_i(\gamma_i)}$ for the convex function $C_i(\cdot)$ where $\overline{\gamma_i}\triangleq \frac{1}{T_o}\sum_{t=0}^{T_o-1}\gamma_{i,t}$ for $i=u,d$,
from \eqref{eqn:T slot drift plus penalty, 2}, we have
\begin{align}\label{eqn:T slot drift plus penalty, 3}
&\left(\frac{1}{T_o}\sum_{t=0}^{T_o-1}E_tP_t\right)+\overline{x_e}+ C_u(\overline{\gamma_u})+\alpha C_d(\overline{\gamma_d})\nn\\
&\leq
\frac{1}{T_o}\sum_{t=0}^{T_o-1}\left[E_t P_t+x_{e,t}+C_u(\gamma_{u,t})+\alpha C_d(\gamma_{d,t})\right]
\end{align}
For a continuously differentiable convex function $f(\cdot)$, it satisfies $f(x)\geq f(y)+f'(y)(x-y)$.
Applying this to $C_u(\overline{x_u})$ and $C_u(\overline{\gamma_u})$, we have
\begin{align}\label{eqn:app3 convex2}
C_u(\overline{x_u})
&\leq C_u(\overline{\gamma_u})+C_u'(\overline{x_u})(\overline{x_u}-\overline{\gamma_u})\nn\\
&\leq C_u(\overline{\gamma_u})+C_u'(\Gamma_u)(\overline{x_u}-\overline{\gamma_u})\nn\\
&=C_u(\overline{\gamma_u})-C_u'(\Gamma_u)\frac{H_{u,T_o}-H_{u,0}}{T_o}
\end{align}
Similarly, we have
\begin{align}\label{eqn:app3 convex3}
C_d(\overline{d_w})\leq C_d(\overline{\gamma_d})-C_d'(\Gamma_d)\frac{H_{d,T_o}-H_{d,0}}{T_o}.
\end{align}
Apply the inequalities in \eqref{eqn:app3 convex2} and \eqref{eqn:app3 convex3} to $C_u(\overline{\gamma_u})$ and $C_d(\overline{\gamma_d})$ respectively at  LHS of \eqref{eqn:T slot drift plus penalty, 3}, and combining \eqref{eqn:T slot drift plus penalty, 2} and \eqref{eqn:T slot drift plus penalty, 3}, we have the  bound of the objective value $u^*(V)$ of {\bf P1} in \eqref{thm1:bd} achieved by our proposed algorithm.

To show the bound in \eqref{thm1:bd_longterm}, as $T_o\to \infty$, it is suffice to show that both $H_{u,t}$ and $H_{d,t}$ in \eqref{thm1:bd} are bounded.
To show these bounds, we need to show that the one-slot Lyapunov drift in \eqref{eqn:drift part1} is upper bounded as follows
\begin{align}\label{eqn:one slot drift bound G}
L(\Thetabf_{t+1})-L(\Thetabf_{t})\leq G.
\end{align}
To show the above bound for the drift, we choose an alternative feasible solution $\tilde{\pibf}_t$ satisfying the following per slot relations:
i) $\tilde{Q}_t+\tilde{S}_{r,t}=\tilde{D}_t+\frac{\Delta_u}{T_o}$;
ii) $\tilde{\gamma}_{i,t}=\tilde{x}_{i,t},\ \textrm{for}\  i=u,d$; and iii) $\tilde{d}_{t}\leq d^{\max}$.
With these relations, and by choosing $\tilde{d}_{t}=d^{\max}$, the terms at RHS of \eqref{eqn:drift z}, \eqref{eqn:Hu part1}, \eqref{eqn: X upper bound} and \eqref{eqn:H2 upper bound} become zeros, and we have \eqref{eqn:one slot drift bound G}.
Averaging \eqref{eqn:one slot drift bound G} over $T_o$-slot period, we have
$\frac{1}{T_o}\left[L(\Thetabf_{T_o})-L(\Thetabf_{0})\right]\leq G$. For any initial value of  $L(\Thetabf_{0})<+\infty$, by \eqref{eqn:lyapunov function}, we have
\begin{align}\label{eqn:avg drift bound G 2}
\hspace*{-1em}\frac{1}{2T_o}\left[Z_{T_o}^2+ H_{u,T_o}^2+\mu\left(X_{T_o}^2+H_{d,T_o}^2\right)\right]
\leq G+\frac{L(\Thetabf_{0})}{T_o}.
\end{align}
It follows that $H_{u,T_o}
\leq \sqrt{2T_oG+2L(\Thetabf_{0})}$ and $H_{d,T_o}
\leq \sqrt{(2T_oG+2L(\Thetabf_{0}))/\mu}$.
Since ${\sqrt{2T_oG+2L(\Thetabf_{0})}}/{T_o}\rightarrow0$ as $T_o\to \infty$, for any initial values of $H_{u,0}$ and $H_{d,0}$,
the third term in RHS of \eqref{thm1:bd} goes to zero.
Thus, we have \eqref{thm1:bd_longterm}.
\endIEEEproof

\section{Proof of Proposition~\ref{prop4}}\label{appE}
\IEEEproof To prove  $\epsilon_d$ is bounded, we note that
\begin{align}\label{eqn:abs epsilon bound}
|\epsilon_d|=\frac{\left|X_{T_o}-X_0\right|}{T_0}
\leq \frac{\left|X_{T_o}\right|+\left|X_{0}\right|}{T_0}.
\end{align}
From \eqref{eqn:avg drift bound G 2}, it follows that
$|X_{T_o}|
\leq \sqrt{(2T_oG+2L(\Thetabf_{0}))/\mu}$.
Substituting the above upper bound of $|X_{T_o}|$ in \eqref{eqn:abs epsilon bound}, we have \eqref{eqn:error dmax}.
\endIEEEproof

\bibliographystyle{IEEEtran}

\begin{IEEEbiography}[{\includegraphics[width=1in,height=1.25in,clip,keepaspectratio]{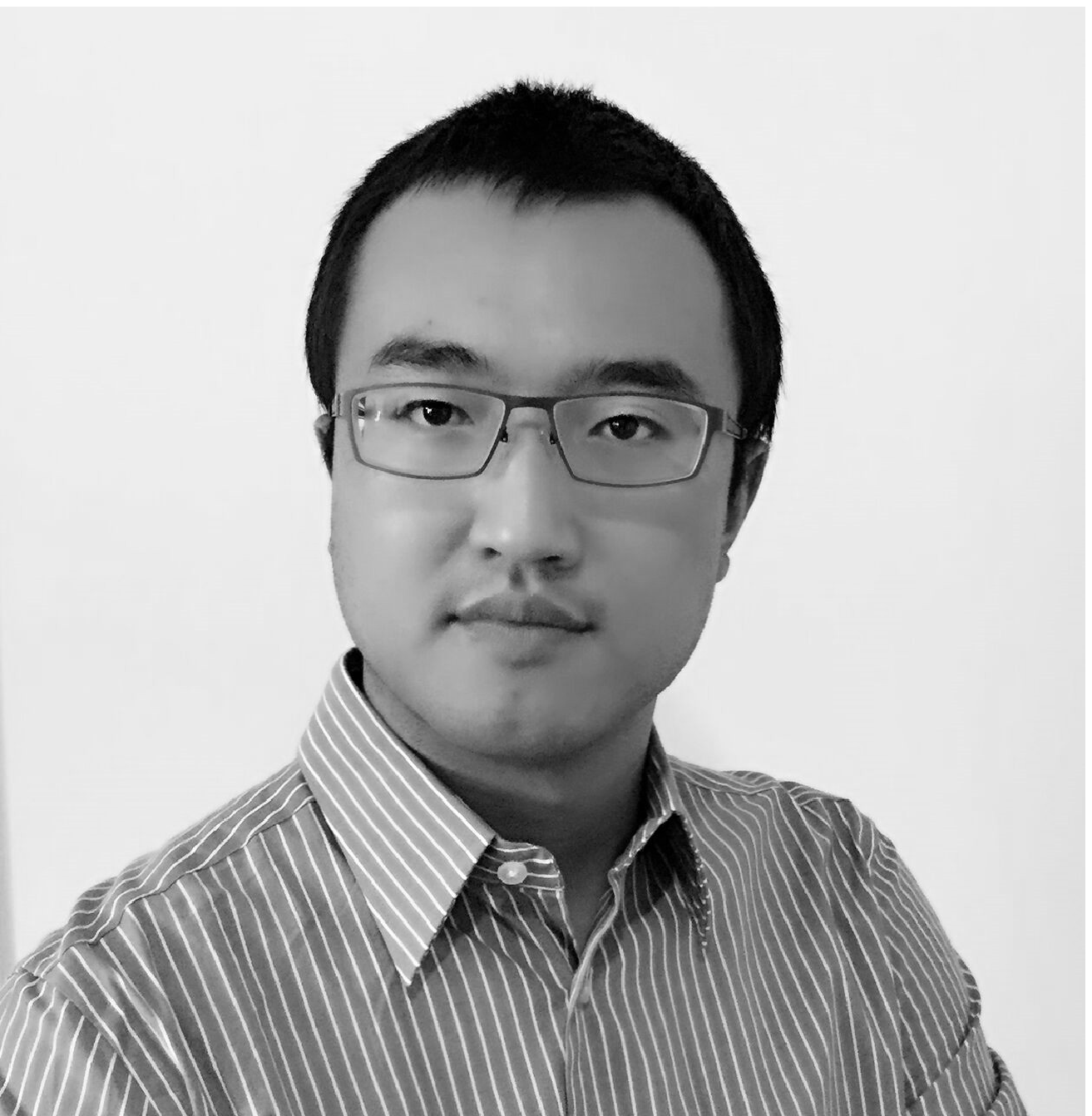}}]{Tianyi Li}
(S'11) received the B.S degrees in communication engineering from North China University of Technology, Beijing, China, and in telecommunication engineering technology from Southern Polytechnic State University, Marietta, GA, USA both in 2009, and the M.Sc degree in electrical engineering from Northwestern University, Evanston, IL, USA in 2011. He obtained his Ph.D degree in 2015 from the Department of Electrical, Computer and Software Engineering, University of Ontario Institute of Technology, Oshawa, Ontario, Canada, and currently is a research associate in the same university. 
He received the best student paper in the Technical Track of Signal Processing for Communications and Networking (SP-COM) at IEEE ICASSP in 2016.
His research interests include stochastic optimization and control in energy storage management and demand side management of smart grid.

\end{IEEEbiography}

\begin{IEEEbiography}[{\includegraphics[width=1in,height=1.25in,clip,keepaspectratio]{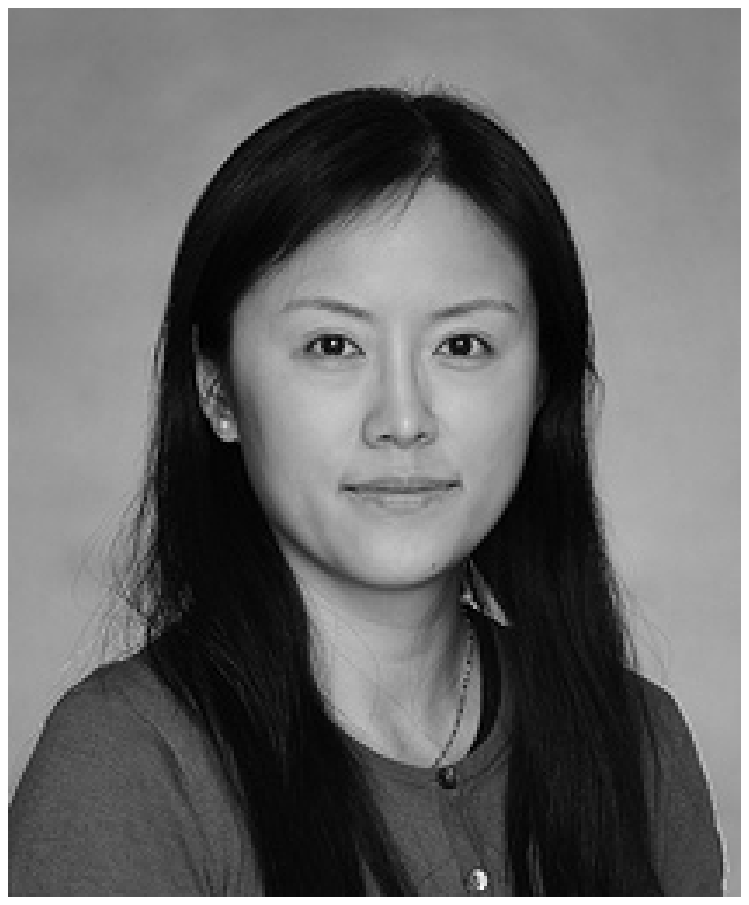}}]{Min Dong}
(S'00-M'05-SM'09) received the B.Eng. degree from Tsinghua University, Beijing, China, in 1998, and the Ph.D. degree in electrical and computer engineering with minor in applied mathematics from Cornell University, Ithaca, NY, in 2004. From 2004 to 2008, she was with Corporate Research and Development, Qualcomm Inc., San Diego, CA. In 2008, she joined the Department of Electrical, Computer and Software Engineering at University of Ontario Institute of Technology, Ontario, Canada, where she is currently an Associate Professor. She also holds a status-only Associate Professor appointment with the Department of Electrical and Computer Engineering at University of Toronto. Her research interests are in the areas of statistical signal processing for communication networks, cooperative communications and networking techniques, and stochastic network optimization in dynamic networks and systems.

Dr. Dong received the Early Researcher Award from Ontario Ministry of Research and Innovation in 2012, the Best Paper Award at IEEE ICCC in 2012, and the 2004 IEEE Signal Processing Society Best Paper Award. She is a co-author of the best student paper in the Technical Track of Signal Processing for Communications and Networking (SP-COM) at IEEE ICASSP in 2016. She served as an Associate Editor for the IEEE TRANSACTIONS ON SIGNAL PROCESSING (2010-2014), and as an Associate Editor for the IEEE SIGNAL PROCESSING LETTERS (2009-2013). She was a symposium lead co-chair of the Communications and Networks to Enable the Smart Grid Symposium at the IEEE International Conference on Smart Grid Communications (SmartGridComm) in 2014. She has been an elected member of IEEE Signal Processing Society Signal Processing for Communications and Networking (SP-COM) Technical Committee since 2013.
\end{IEEEbiography}
\end{document}